\documentclass[traditabstract]{aa} 

\usepackage{amssymb}
\usepackage{amsmath}
\usepackage{hyperref}
\usepackage{graphicx,array,txfonts}
\usepackage{natbib}
\usepackage{booktabs}
\usepackage{caption}
\usepackage{subcaption}
\usepackage{placeins}
\usepackage[toc,page]{appendix} 
\usepackage{gensymb}

\usepackage[export]{adjustbox}

\usepackage{comment}
\usepackage{threeparttable}
\usepackage{multicol}
\usepackage{tabularx}
\usepackage{lscape}
\def\GLEE{\textsc{Glee}}

\usepackage{color}
\usepackage{xcolor}
\usepackage{longtable}

\usepackage{orcidlink}
\usepackage{float}

\usepackage[normalem]{ulem}

\begin{document}

\title{Cosmology with supernova Encore in the strong lensing cluster MACS J0138$-$2155}
\subtitle{Photometry, cluster members, and lens mass model}

\titlerunning{Cosmology with supernova Encore}

\author{S.~Ertl\inst{\ref{mpa},\ref{tum}} \and
        S.~H.~Suyu\inst{\ref{tum},\ref{mpa}}\orcidlink{0000-0001-5568-6052} \and
        S.~Schuldt\inst{\ref{unimi},\ref{inafmilano}}\orcidlink{0000-0003-2497-6334} \and
        G.~Granata\inst{\ref{unimi},\ref{uferrara},\ref{icg}}\orcidlink{0000-0002-9512-3788} \and
        C.~Grillo\inst{\ref{unimi},\ref{inafmilano}}\orcidlink{0000-0002-5926-7143} \and
        G.~B.~Caminha\inst{\ref{tum},\ref{mpa}} \and   
        A.~Acebron\inst{\ref{unican},\ref{inafmilano}} 
        \and
        P.~Bergamini\inst{\ref{unimi},\ref{inafbologna}} \and
        R.~Ca\~nameras\inst{\ref{aix}} \and
        S.~Cha\inst{\ref{yonsei}}\orcidlink{0000-0001-7148-6915}
        \and
        J.~M.~Diego\inst{\ref{unican}} \and
        N.~Foo\inst{\ref{arizona}} \and
        B.~L.~Frye\inst{\ref{arizona}} \and
        Y.~Fudamoto\inst{\ref{chiba},\ref{arizona}}\orcidlink{0000-0001-7440-8832} \and
        A.~Halkola\inst{\ref{tuusula}} \and
        M.~J.~Jee\inst{\ref{yonsei},\ref{ucd}}\orcidlink{0000-0002-5751-3697}
        \and
        P.~S.~Kamieneski\inst{\ref{asu}} \and{A.~M.~Koekemoer}\inst{\ref{stsci}}\orcidlink{0000-0002-6610-2048} \and
        {A.~K.~Meena}\inst{\ref{BenGurion}}\orcidlink{0000-0002-7876-4321} \and
        S.~Nishida\inst{\ref{chibaphys}} \and
        M.~Oguri\inst{\ref{chiba},\ref{chibaphys}} \and
        J.~D.~R.~Pierel\inst{\ref{stsci},\ref{ef}}\orcidlink{0000-0002-2361-7201} \and
        P.~Rosati\inst{\ref{uferrara},\ref{inafbologna}}\orcidlink{0000-0002-6813-0632} \and
        {L.~Tortorelli}\inst{\ref{LMU}} \and
        H.~Wang\inst{\ref{mpa},\ref{tum}} \and {A.~Zitrin}\inst{\ref{BenGurion}}\orcidlink{0000-0002-0350-4488}
}

\institute{
    Max-Planck-Institut f{\"u}r Astrophysik, Karl-Schwarzschild Stra{\ss}e 1, 85748 Garching, Germany\\
    e-mail: \href{mailto:ertlseb@mpa-garching.mpg.de}{\tt ertlseb@mpa-garching.mpg.de} \label{mpa}
    \and
    Technical University of Munich, TUM School of Natural Sciences, Physics Department,  James-Franck-Stra{\ss}e 1, 85748 Garching, Germany \label{tum}
    \and 
    Dipartimento di Fisica, Universit\`a  degli Studi di Milano, via Celoria 16, I-20133 Milano, Italy
    \label{unimi}
    \and 
    INAF -- IASF Milano, via A.~Corti 12, I-20133 Milano, Italy
    \label{inafmilano}
    \and
    Dipartimento di Fisica e Scienze della Terra, Università degli Studi di Ferrara, via Saragat 1, 44122 Ferrara, Italy\label{uferrara}
    \and
    Institute of Cosmology and Gravitation, University of Portsmouth, Burnaby Rd, Portsmouth PO1 3FX, UK \label{icg}
    \and
    Instituto de F\'isica de Cantabria (CSIC-UC), Avda.~Los Castros s/n, 39005 Santander, Spain 
    \label{unican}
    \and
    INAF -- OAS, Osservatorio di Astrofisica e Scienza dello Spazio di Bologna, via Gobetti 93/3, I-40129 Bologna, Italy
    \label{inafbologna}
    \and
    Aix-Marseille Université, CNRS, CNES, LAM, Marseille, France
    \label{aix}
    \and
    Department of Astronomy, Yonsei University, 50 Yonsei-ro, Seoul 03722, Korea
    \label{yonsei}
    \and
    Department of Astronomy/Steward Observatory,
    University of Arizona,
    933 N. Cherry Avenue,
    Tucson, AZ 85721, USA
    \label{arizona}
    \and
    Center for Frontier Science, Chiba University, 1-33 Yayoi-cho, Inage-ku, Chiba 263-8522, Japan
    \label{chiba}
    \and
    Py\"orrekuja 5 A, 04300 Tuusula, Finland
    \label{tuusula}
    \and
    Department of Physics and Astronomy, University of California, Davis, One Shields Avenue, Davis, CA 95616, USA
    \label{ucd}
    \and
    School of Earth and Space Exploration, Arizona State University, PO Box 876004, Tempe, AZ 85287-6004, USA
    \label{asu}
    \and
    Space Telescope Science Institute,
    3700 San Martin Drive, Baltimore, MD 21218, USA
    \label{stsci}
    \and
    Department of Physics, Ben-Gurion University of the Negev, P.O. Box 653, Be'er-Sheva 84105, Israel
    \label{BenGurion}
    \and
    Department of Physics, Graduate School of Science, Chiba University, 1-33 Yayoi-Cho, Inage-Ku, Chiba 263-8522, Japan 
    \label{chibaphys}
    \and
    University Observatory, Ludwig-Maximilians-Universit\"at M\"unchen, 
    Scheinerstraße 1, 81679, M\"unchen, Germany
    \label{LMU}
    \and
    NASA Einstein Fellow
    \label{ef}
}

\date{Received 11 March 2025; accepted 13 July 2025}

\abstract 
{The strongly lensed supernova (SN) Encore, at a redshift of $z = 1.949$ and discovered behind the galaxy cluster MACS J0138$-$2155 at $z=0.336$, provides a rare opportunity for time-delay cosmography and studies of the SN host galaxy, where previously another SN, called SN Requiem, had appeared. To enable these studies, we combined new James Webb Space Telescope (JWST) imaging, archival Hubble Space Telescope (HST) imaging, and new Very Large Telescope (VLT) spectroscopic data to construct state-of-the-art lens mass models that are composed of cluster dark-matter (DM) haloes and galaxies.  We fitted the surface brightness distributions of the galaxies in the field of view using S{\'e}rsic profiles to determine their photometric and structural parameters across six JWST and five HST filters.  We used the colour-magnitude and colour-colour relations of spectroscopically confirmed cluster members to select additional cluster members, and identified a total of 84 galaxies belonging to the galaxy cluster.  We constructed seven different mass models using a variety of DM halo mass profiles and explored both multi-plane and approximate single-plane lens models.  As constraints, we used the observed positions of 23 multiple images from eight multiple image systems that originate from four galaxies with distinct spectroscopic redshifts in the range of 0.767 to 3.420.  In addition, we used stellar velocity dispersion measurements to obtain priors on the galaxy mass distributions.  We find that six of the seven models fit well to the observed image positions, with a root-mean-square (rms) scatter of $\leq0.032\arcsec$ between the model-predicted and observed positions for systems identified with JWST and HST images, including SN Encore and SN Requiem (the rms scatter is $0.24\arcsec$ for all positions, including those identified with MUSE images). Mass models with cored-isothermal DM profiles fit well to the observations, whereas the mass model with a Navarro-Frenk-White cluster DM profile has an image-position $\chi^2$ value that is four times higher.  We built our ultimate model by combining four multi-lens-plane mass models in order to incorporate uncertainties due to model parameterizations. Our two approximate mass models with a single-lens plane allow us to perform direct comparisons with single-plane models built independently by other teams.  Using our ultimate model, we predict the image positions and magnifications of SN Encore and SN Requiem.  We also provide the effective convergence and shear of SN Encore for micro-lensing studies.  Our work lays the foundation for building state-of-the-art mass models of the cluster for future cosmological analysis and SN host galaxy studies. 
}


\maketitle
%
\section{Introduction}
\label{sec:intro}

Strong gravitational lensing by galaxy clusters can be used as a powerful and versatile tool in astrophysics and cosmology. For instance, the sensitivity of gravitational lensing to the total mass distribution of an object allows us to probe its dark matter (DM) content \citep[e.g.][]{Grillo2015,schuldt19,Limousin2022,Wang2022,ChaJee2023}. Properties of high redshift galaxies that would otherwise be undetected are revealed thanks to the lensing magnification effect that turns gravitational lenses into `cosmic telescopes' \citep[e.g.][]{Coe2012,Alavi2016,Acebron2018,Bouwens2022,Castellano2023}.

In recent years, supernovae (SNe) strongly lensed by galaxy clusters have gained importance as a probe of cosmological parameters such as the Hubble constant, $H_0$, which is the current expansion rate of the Universe. Its value has been measured from the cosmic microwave background with the Planck satellite \citep[][$H_0=67.4 \pm 0.5 \,\textrm{km s}^{-1} \textrm{Mpc}^{-1}$]{Aghanim2020}, and through the type Ia SN distance ladder calibrated with Cepheid \citep[][$H_0=73.0 \pm 1.0\, \textrm{km s}^{-1} \textrm{Mpc}^{-1}$]{Riess2022} or in combination with stars from the tip of the red giant branch and J-region asymptotic giant branch stars \citep[][$H_0 = 69.96 \pm 1.05 {\rm \, (stat)} \pm 1.12 {\rm \, (sys)} \, \textrm{km s}^{-1} \textrm{Mpc}^{-1}$]{Freedman2024}. The increased precision of these measurements from \citet{Aghanim2020} and \citet{Riess2022} has led to a $\sim 5 \sigma$ tension. This tension could reveal unknown systematic effects in the measurements \citep[e.g.][]{Efstathiou2020,Yeung2022,Freedman2023,Riess2024} or could be evidence for new physics beyond the standard flat $\Lambda$CDM \citep{diValentino2021} model, which is a spatially flat universe consisting of dark energy described by the cosmological constant $\Lambda$ and cold dark matter (CDM).

An independent way of measuring $H_0$ is through time-delay cosmography. This method, first proposed by \citet{Refsdal1964}, makes use of the variability of objects such as quasars or SNe, which are observed multiple times through the strong lensing effect. Because of the different path lengths the light takes to get to the multiple image positions and the differences in the gravitational potential, the variability will be observed with a time delay between the images. The time delays are measured from light curves obtained from monitoring by the COSmological MOnitoring of GRAvItational Lenses \citep[COSMOGRAIL;][]{Courbin2004,Vuissoz2008,Courbin2011,Tewes2013, Millon2020a, Millon2020b} collaboration for lensed quasars, for example. The time delays are related to the lensing potential, $\psi,$ of the lens, and the so-called time-delay distance, $D_{\Delta t}$, which is proportional to $1/H_0$ \citep{Suyu2010}. 

The method of time-delay cosmography has been applied successfully to lensed quasars \citep[e.g.][]{Suyu2010,Suyu2014,Wong2017,Jee2019,Birrer2019,Chen2019,Rusu2020, Shajib2020, Wong2024} in the $H_0$ lenses in COSMOGRAIL's Wellspring \citep[H0LiCOW;][]{Suyu2017}, COSMOGRAIL, STRong lensing Insights into the Dark Energy Survey \citep[STRIDES;][]{Treu2018}, and the Strong lensing at High Angular Resolution Program \citep[SHARP;][]{Lagattuta2012,Chen2022} collaborations. The $H_0$ inferences of six individual lenses were combined by the H0LiCOW collaboration for a joint constraint on $H_0$, yielding $H_0=73.3_{-1.8}^{+1.7} \textrm{km s}^{-1} \textrm{Mpc}^{-1}$ \citep{Wong2020}. The Time-Delay COSMOgraphy (TDCOSMO) collaboration showed that relaxing the assumptions on the mass model and using only stellar velocity dispersion measurements of the lens galaxies to constrain the shape of the mass profile results in $H_0=74.5_{-6.1}^{+5.6} \textrm{km s}^{-1} \textrm{Mpc}^{-1}$, i.e. the uncertainty on $H_0$ increases from 2\% to 8\% but the median value does not change significantly \citep{Birrer2020}.

Lensed SNe can be used in a similar way to determine the Hubble constant. The advantages of SNe over quasars for $H_0$ inference are that their light curves change drastically on timescales of weeks, thereby providing precise time delays with less monitoring \citep{Pierel2019,Pierel2021, Huber+2022, HuberSuyu2024}. Furthermore, SNe, in contrast to quasars, will fade away, making the modelling of the light distribution of the lens galaxy or the arc more precise \citep{Ding2021}. In the case of (at least) SNe Ia, chromatic microlensing effects (due to stars and compact objects in the lensing galaxy) on the light curves are mitigated in the early parts of colour curves when microlensing is achromatic due to similar spatial intensity distributions across wavelengths in the early phases of SNe \citep[e.g.][]{Goldstein2018,Huber2021}. For sources lensed by galaxy clusters, the time delays are of the order of months to years and can be measured with precisions of 1-3\% \citep{Fohlmeister2013,Dahle2015,Kelly2016b,Kelly2023b}. 
From the eight lensed SN systems discovered so far (six cluster-scale lenses and two galaxy-scale lenses), two had sufficiently long time-delay measurements for time-delay cosmography, both lensed by galaxy clusters. 

The first $H_0$ measurement from a lensed SN came from images of SN Refsdal \citep{Kelly2015,Kelly2016a} at $z=1.49$ in the Hubble Frontier Field cluster MACS J1149.5+2223. It was classified as a type II SN based on the shape of the light curve and from spectroscopy \citep{Kelly2016a}. With the measured time delays and combination of eight lens models, \citet{Kelly2023} inferred $H_0=64.8_{-4.3}^{+4.4} \, \textrm{km s}^{-1} \textrm{Mpc}^{-1}$, and $H_0=66.6_{-3.3}^{+4.1} \,\textrm{km s}^{-1} \textrm{Mpc}^{-1}$ using the two models most consistent with the observations, which is roughly a 6\% precision on $H_0$. By strictly following their blind modelling methodology presented and frozen in \citet{Grillo2016}, \citet{Grillo2024} measured $H_0=65.1_{-3.4}^{+3.5}\, \textrm{km s}^{-1} \textrm{Mpc}^{-1}$ with SN Refsdal using more relaxed assumptions on the background cosmology. The $H_0$ measurement from \citet{Grillo2024} is robust with respect to cosmological model assumptions as a result of the presence of multiple lensed background sources at different redshifts being lensed by the cluster, in addition to SN Refsdal that provides time-delay constraints.

After SN Refsdal, a SN called `SN H0pe' was detected in the JWST imaging of the PLCK G165.7+67.0 (G165) galaxy cluster field ($z=0.35$), which was triply imaged as a result of strong lensing by G165 \citep[PID 1176;][]{Windhorst2023,Frye2023,Frye2024}.  Follow-up JWST NIRCam and NIRSpec observations (PID 4446) confirmed the SN to be a type Ia at $z=1.783$ \citep{Frye2024,Chen2024,Polletta+23}. Using the three images of SN H0pe, the two relative time delays and three absolute magnifications were measured by \citet{Pierel2024b} using photometry and \citet{Chen2024} using spectroscopy.  The values for these five observables were also predicted by seven different lens modelling approaches all incorporating identical lensing evidence and blinded from each other and from the time-delay measurement team.  A joint fitting of the five predicted observables to the measured values yielded $H_0 = 75.4^{+8.1}_{-5.5}$~km~s$^{-1}$~Mpc$^{-1}$ \citep{Pascale2025}. This is only the second measurement of $H_0$ from a lensed SN after SN Refsdal, and the first using a standardizable candle (a type Ia SN). We refer to the above references, especially \citet{Pascale2025}, for more details on the SN H0pe analysis.
The measurements from SN Refsdal and SN H0pe are consistent with each other at the $\sim1.5-2$$\sigma$ level. 

With the recent discovery of the lensed SN Encore \citep{Pierel2024}, a third system suitable for time-delay cosmography is added to the sample. SN Encore is a type Ia SN at redshift $z=1.95$ that is lensed into multiple images by the galaxy cluster MACS J0138$-$2155 \citep{Newman2018a,Newman2018b}. The same source galaxy was previously host to another likely type Ia SN, called SN Requiem, which was discovered in archival Hubble Space Telescope (HST) data taken in 2016 \citep{Rodney2021}. Since SN Requiem was discovered after the SN had faded, it could not be used to obtain a precise $H_0$ measurement, but \citet{Rodney2021} predicted the appearance of a future image in the year $\sim$2037, the detection of which will yield a time delay with percent-level uncertainty.  In contrast, SN Encore was discovered while the SN was still visible, and \citet{Pierel2024} and \citet{Granata+24} were able to acquire additional observations of SN Encore and the galaxy cluster.  Ongoing time-delay measurements of SN Encore (Pierel et al. in prep.) and the giant lensing arcs of the SN host galaxy provide an excellent opportunity to use this cluster-scale lens as a probe for $H_0$.
In order to infer the Hubble constant, an accurate model of the lensing galaxy cluster MACS J0138$-$2155 is needed in addition to the time delays. Furthermore, the mass model could be used to predict the properties of future images of SN Encore and SN Requiem, which would facilitate their detections. 

In this paper, we present the data processing approach and have collected all the necessary ingredients for building a cluster mass model.  In particular, we measured the photometry of galaxies in the field of the cluster and combined them with newly acquired spectroscopic observations presented in \citet{Granata+24} in order to identify galaxies that are members of the galaxy cluster. This dataset, together with the spectroscopically identified lensed background sources, was shared with seven independent modelling teams using a variety of modelling software.  The cluster mass modelling by each team was performed independently in a blind analysis where the findings of each team were not communicated until all teams finalized their mass models.  The comparison of the mass modelling results from the blind analysis is presented in an upcoming paper (Suyu et al. in prep.).  In this paper, we present the modelling details of the {\sc Glee}\footnote{{\sc Glee} (Gravitational Lens Efficient Explorer) is a lens modelling software \citep{SuyuHalkola2010, Suyu+2012}.} team, which is composed of the first six authors.  The {\sc Glee} model encapsulates the dark matter and galaxy mass components parametrically, and employs multi-lens-plane ray tracing.  In order to maintain a blind analysis throughout our work, the first part of this paper draft (Sect.~\ref{sec:intro} to \ref{sec:selec}), which contains the input data, was shared throughout this work with all modelling teams as the teams checked and all agreed on the same input data for building their mass models. The rest of the paper draft (Sect.~\ref{sec:mass_model} to \ref{sec:summary}) that specifically describes the {\sc Glee} mass model was only accessible to the {\sc Glee} team before unblinding, and was subsequently shared with other teams after unblinding.  The model results from our blind analysis are presented here without further modification. 
The $H_0$ constraint from our mass modelling will be presented in Pierel et al. (in prep.). 

The outline of the paper is as follows.  We present the observational data used in our analysis in Sect.~\ref{sec:obs} and the photometric measurements of galaxies in Sect.~\ref{sec:photometry_sersic}.  Through the spectroscopic data presented in \citet{Granata+24} and the photometric measurements, we identified galaxies that are members of the galaxy cluster, additional galaxies along the line-of-sight that are potentially significant for the cluster mass model, and background source galaxies that are strongly lensed by the cluster in Sect.~\ref{sec:selec}.   The methods and details of our mass modelling are presented in Sect.~\ref{sec:mass_model} and the results, including predictions of SNe Encore and Requiem's positions and magnifications, are in Sect.~\ref{sec:results_fixed_cosmo}. We summarize in Sect.~\ref{sec:summary}. 

Throughout the paper, 
magnitudes are reported in the AB magnitude system and parameter estimates are given by the median of its one-dimensional marginalized posterior probability density function. The quoted uncertainties show the 16th$^{\rm }$ and 84th$^{\rm }$ percentiles (corresponding to a 68\% credible interval).  Since the time-delay measurements are not yet available, our mass models in this work adopt a flat $\Lambda$CDM cosmology with $H_0=70\, \rm{\, km\, s^{-1}\, Mpc^{-1}}$ and $\Omega_{\rm M} = 1 - \Omega_{\Lambda}=0.3$, especially to predict the properties of SN Encore and SN Requiem.

\section{Observations}
\label{sec:obs}

\begin{figure*}
        \centering
        \includegraphics[scale=0.52]{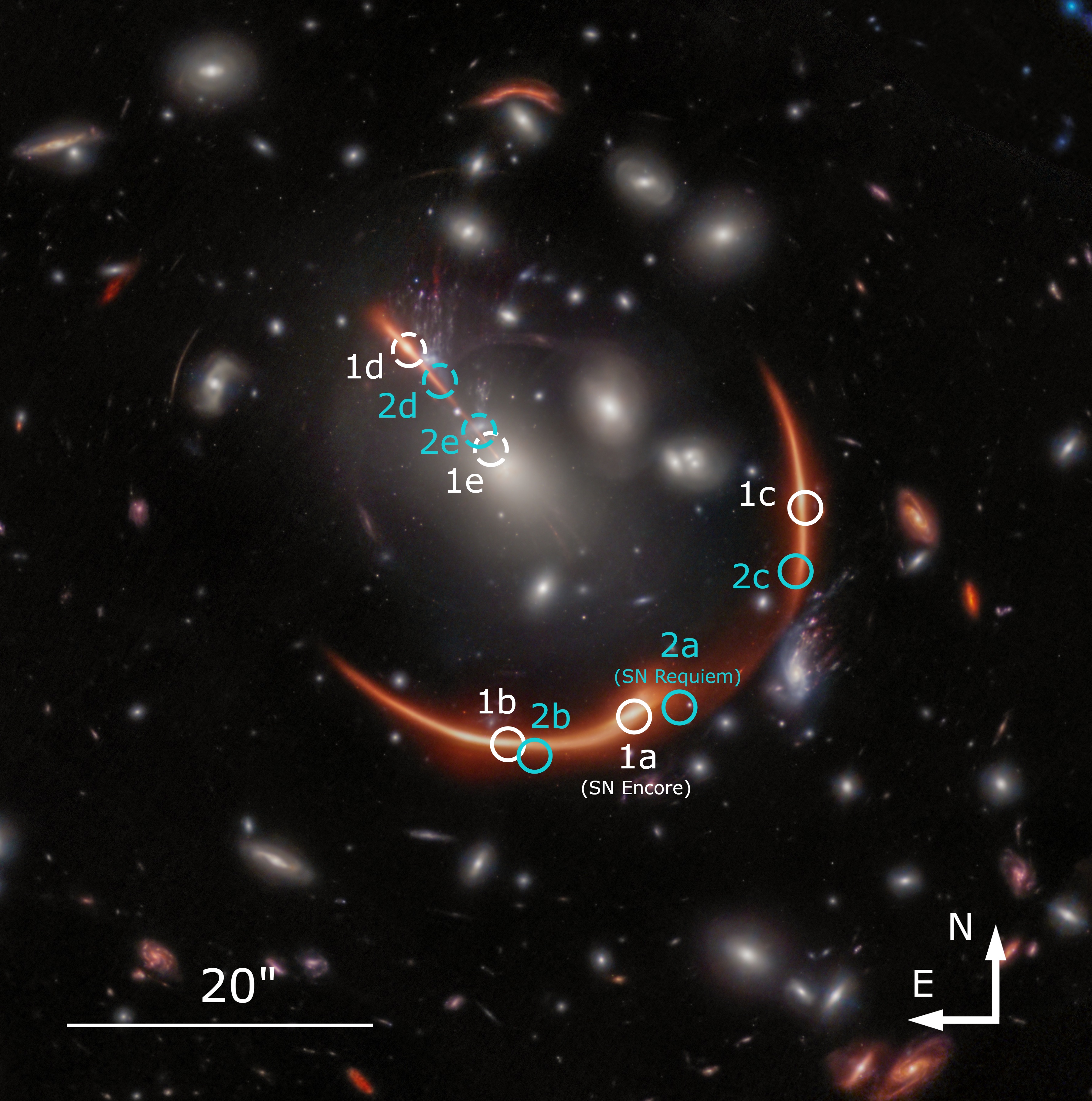}
        \caption{Colour image of MACS J0138$-$2155 from combining JWST/NIRCam and HST/WFC3 data. The RGB image is constructed from the filters F105W+F115W+F125W (blue), F150W+F160W+F200W (green), and F277W+F356W+F444W (red). The observed images of SN Encore are shown in solid white circles, the expected positions of images D and possibly E are in dashed white circles. The observed and predicted images for SN Requiem are shown in blue following the same pattern. The images were drizzled to a pixel scale of 0.04\arcsec/pix. (Image Credit: STScI, A. Koekemoer, T. Li)}
        \label{fig:color_img}
\end{figure*}

MACS J0138$-$2155 (shown in Fig.~ \ref{fig:color_img}) is a galaxy cluster at redshift $z_{\rm d} = 0.336$, discovered within the MAssive Cluster Survey \citep[MACS;][]{Ebeling2001,Repp2018}. This cluster acts as a strong gravitational lens and magnifies the light of a near-infrared (NIR) source galaxy MRG$-$M0138 at a redshift $z_{\rm s} = 1.949$ \citep{Newman2018b} into two tangential arcs and one radial arc, which makes this system one of the brightest extragalactic objects in the NIR \citep{Newman2018a,Newman2018b}. The bright central region of the source galaxy is lensed into at least five multiple images located in the giant tangential arc ($\sim$$20 \arcsec$ wide) to the south of the cluster centre, the tangential arc ($\sim$$15 \arcsec$ wide) west of the brightest cluster galaxy (BCG), and the radial arc north-east of the BCG. The source galaxy is the host of SN Encore \citep{Pierel2024}, which is itself lensed into multiple images. Three images 1a, 1b, and 1c of SN Encore have been observed in the two tangential arcs (see white circles in Fig.~\ref{fig:color_img}, although image 1c has low signal-to-noise ratio and is not clearly visible\footnote{There are hints of its presence in the JWST imaging; a template image of the galaxy cluster after SN Encore will have faded will be beneficial to extract the signal of image 1c.}). A fourth image 1d is expected in the future in the radial arc, and a possible central image 1e (fifth image, also in the future with the longest time delay) would be in the central region of the bright BCG (dashed white circles in Fig.~\ref{fig:color_img}). To build the total cluster mass model we use imaging data from the Hubble Space Telescope (HST) and the James Webb Space Telescope (JWST), as well as spectroscopic observations from Multi Unit Spectroscopic Explorer \citep[MUSE;][]{Bacon2010}, which are described in this section.

\subsection{HST and JWST imaging}
\label{sec:obs:HST_JWST}

\newcolumntype{L}{>{\raggedright\arraybackslash}X}

\begin{table*}[h]
\caption{HST observations.}
\centering
\begin{tabularx}{\linewidth}{p{3cm}p{3cm} *{8}{L}}\toprule \toprule
        Program  & Dates & \multicolumn{8}{c}{Filters} \\ \midrule \midrule
        14496 (PI: Newman)    & 2016 Jun 03 &  & F555W & &  &  &  & &    \\
                             & 2016 Jul 18/19  & & & & F105W &  &  & & F160W    \\
        15663 (PI: Akhshik)  & 2019 Jul 13  &  & & & & F110W & &  &    \\ 
                              & 2019 Jul 14 &  & & & & & & F140W &   \\ 
                            & 2019 Jul 19 & & & &  & & F125W & &  \\ 
                            & 2019 Jul 21 & F390W  &  & F814W  &  &  &  & F140W &   \\

        16264 (PI: Pierel)  & 2023 Dec 16 & & & & F105W &  & F125W  & & F160W   \\
                             & 2024 Jan 30  & & & & F105W &  & F125W  & & F160W    \\
             \toprule  
            
                \ \
        \label{tab:encore_hst_obs}
\end{tabularx}
\end{table*}

\begin{table*}[h]
\caption{Summary of JWST observations.}
\centering
\begin{tabularx}{\linewidth}{p{3cm}p{2.5cm} *{6}{L}}\toprule \toprule
        Program  & Dates & \multicolumn{6}{c}{Filters} \\ \midrule \midrule
        2345 (PI: Newman)   & 2023 Nov 17  &  & F150W &  &  &  & F444W   \\
        6549 (PI: Pierel)     & 2023 Dec 5  & F115W & F150W & F200W & F277W & F356W & F444W  \\ 
            & 2023 Dec 23  & F115W & F150W & F200W & F277W & F356W & F444W   \\
        & 2024 Jan 8 & F115W & F150W & F200W & F277W & F356W & F444W   \\
             \toprule  
            
                \ \

\end{tabularx}
\label{tab:encore_jwst_obs}
\end{table*}

The first visit of HST was in 2016 (Proposal ID 14496, PI: Newman) when ACS and WFC3 images were taken. In the scope of the REQUIEM galaxy survey \citep{Akhshik2020,Akhshik2023}, 6 orbits of WFC3 observations were performed in 2019 (Proposal ID  15663, PI: Akhshik). SN Encore was discovered in JWST data taken in November 2023 (Proposal ID 2345, PI: Newman), which led to follow-up observations in December 2023 and January 2024 with HST (Proposal ID 16264, PI: Pierel) and JWST (Proposal ID 6549, PI: Pierel). We summarize the HST and JWST observations in Table~\ref{tab:encore_hst_obs} and Table~\ref{tab:encore_jwst_obs}, respectively. The newly obtained HST and JWST imaging data and reduction are described in detail in \citet{Pierel2024}. 

All available imaging data were combined into mosaics with an updated version of the HST 'mosaicdrizzle' pipeline \citep{Koekemoer2011}. This produced separate epoch mosaics, as well as full depth mosaics where all epochs per filter were combined. Each mosaic product contains multiple extensions, including the science image and the error image (providing the 1$\sigma$ uncertainties of the science image pixels).  The absolute astrometry of all mosaics is directly aligned to Gaia-DR3 \citep{Prusti2016,Vallenari2023}. In Table~\ref{tab:encore_img_data}, we summarize all imaging data products that are used in our analysis. The filters were chosen such that a large wavelength range is covered with preference for filters with a large total exposure time. The final data products are drizzled to a pixel size of 0.04\arcsec / pixel for all the filters in order to balance between the different native pixel sizes of the different detectors and to sample sufficiently the point-spread functions (PSFs).

\begin{table}[h]
\caption{Full-depth imaging data used in this work.}
\centering
    \begin{tabular}{lp{2.5cm}cc}\toprule
        Telescope & Instrument/Filter & $m_{\rm 5\sigma}$ & $t_{\rm exp}$ [s] \\ \midrule \midrule \vspace{5px} 
        HST & ACS/F555W  & 27.938 & \phantom{1}5214 \\\vspace{5px} 
             & WFC3/F814W  & 26.441 & \phantom{11}912  \\ \vspace{5px} 
             & WFC3/F105W  & 27.870 & 17438  \\ \vspace{5px} 
             & WFC3/F125W  & 27.302 & \phantom{1}9923  \\ \vspace{5px} 
            & WFC3/F160W  & 27.297 & \phantom{1}8525  \\ \vspace{5px} 
        JWST & NIRCam/F115W  & 27.196 & \phantom{1}4467   \\ \vspace{5px} 
             & NIRCam/F150W  & 27.156 & \phantom{1}3693  \\ \vspace{5px} 
            & NIRCam/F200W  & 27.246 & \phantom{1}2920  \\ \vspace{5px} 
            & NIRCam/F277W  & 27.921 & \phantom{1}2920  \\ \vspace{5px} 
            & NIRCam/F356W  & 28.074 & \phantom{1}2920  \\ \vspace{5px} 
            & NIRCam/F444W  & 28.087 & \phantom{1}5240  \\
             \toprule  
                \end{tabular}
            
                \ \
        \caption*{\textbf{Notes: } We use the full depth data drizzled to a scale of 0$\farcs$04/pix. The first two columns specify the telescope and instruments, the third column, $m_{\rm 5\sigma}$ , is the 5$\sigma$ limiting magnitude depth (AB) determined from 0$\farcs$15 radius apertures in empty regions across the mosaics, and the fourth column, $t_{\rm exp}$ , is the total exposure time depth of the mosaics.}
        \label{tab:encore_img_data}
\end{table}

\subsection{MUSE spectroscopy and redshift measurements}
\label{sec:obs:muse}

MACS J0138$-$2155 was the target of two programs carried out by the integral-field spectrograph MUSE \citep{Bacon2010}, at the Very Large Telescope of the European Southern Observatory. The first $\sim$$49$~minutes were obtained within the program ID 0103.A-0777 (September 2019, PI: Edge), later complemented by additional 2.9 hours of data collected after the detection of SN Encore in the Target of Opportunity program ID 110.23PS (December 2023, PI: Suyu). The average seeing across the observations was approximately $0.8''$. The data reduction pipeline adopted for the creation of the final data-cube is described in detail in \citet{Granata+24}.

The MUSE spectroscopy was used in \citet{Granata+24} as the basis for the construction of a redshift catalogue containing 107 reliable redshift measurements of extragalactic objects. They identified 50 cluster members in the redshift range $z=0.324-0.349$, that yielded a cluster redshift of $z_{\rm d}=0.336$. Finally, they performed a blind search of spectral features appearing simultaneously at different locations across the data cube, which allowed for the identification of two additional background sources systems 5 and 6 with secure redshifts that are strongly lensed by the galaxy cluster, and an additional source with a tentative redshift measurement that is strongly lensed by a cluster member (this source is not included in our final model).  In total, there are four independent background source galaxies with secure spectroscopic redshifts from $z = 0.767$ to $3.420$ that are strongly lensed into multiple image systems.

\section{Photometry and S\'{e}rsic profiles for objects in MACS J0138$-$2155}
\label{sec:photometry_sersic}
Accurate photometry of galaxies in the cluster field of view is important to photometrically select cluster members for the mass model, as well as to scale their mass parameters with a scaling relation for the modelling. Therefore, we fitted the surface brightness of several objects in the field using S\'{e}rsic profiles \citep{Sersic.1963} described in Sect.~\ref{sec:photometry_sersic:sersic}.  The S\'{e}rsic profile fitting gives accurate photometry, especially in crowded fields of galaxy clusters. We subtracted the extended light of the BCG to avoid biases in the photometry of nearby galaxies. This procedure is described in Sect.~\ref{sec:photometry_sersic:BCG}, and the photometry of other objects in the field in Sect.~\ref{sec:photometry_sersic:Sersic_fit}.

\subsection{S\'{e}rsic profiles}
\label{sec:photometry_sersic:sersic}

To measure the magnitudes of the galaxies, we fitted their two-dimensional surface brightness (SB) with S\'{e}rsic profiles \citep{Sersic.1963} in all filters listed in Table~\ref{tab:encore_img_data}. The S\'{e}rsic profile is parametrized as
\begin{equation}
        I_{\rm S}(x,y) = A_{\rm S}\exp\Bigg[-k\Bigg\{\Bigg(\frac{\sqrt{(x-x_{\rm S})^2+\left( \frac{y-y_{\rm S} }{q_\text{S}} \right) ^2}}{r_{\rm eff}}\Bigg)^{\frac{1}{n_{\rm s}}}-1\Bigg\}\Bigg],
        \label{eq:sersic}
\end{equation}
where $A_{\rm S}$ is the amplitude, $x_{\rm S}$ and $y_{\rm S}$ are the centroid coordinates, $q_{\rm S}$ is the axis ratio ($0<q_{\rm S}\leq1$), and $n_{\rm S}$ the S\'{e}rsic index. The value for the normalization constant $k$ is set such that $r_{\rm eff}$ is the half-light radius in the direction of the semi-major axis ($k$ is thus not a freely variable parameter). The light distribution is rotated by a position angle $\phi_{\rm S}$, which is measured east of north (where $\phi_{\rm S}=0^\circ$ corresponds to the major axis being aligned with the northern direction).

We constructed the point-spread function (PSF) from multiple stars in the field with the software \texttt{STARRED} \citep{Michalewicz2023,Millon2024}. We manually chose bright, unsaturated stars in the field. The number of stars used depends on the field of view (FOV), but we used at least three stars for each filter. 

The galaxy SB can be modelled as a combination of multiple  S\'{e}rsic components to obtain SB distribution that fit better to the observations.  In that case, the total SB is 
\begin{equation}
        I_{\rm S,tot}(x,y) = \sum_{i=1}^{N_{\rm sersic}}
        I_{{\rm S}, i}(x,y),
        \label{eq:Isersicsum}
\end{equation}
where $N_{\rm sersic}$ is the number of S\'{e}rsic  components. 

The S\'{e}rsic parameters are optimized to fit to the observed intensity value $I_i^{\rm obs}$ of pixel $i$ by minimizing the $\chi^2$ function of the SB
\begin{equation}
        \chi^2_{\rm SB} = \sum_{i=1}^{N_{\rm p}}\frac{|I_i^{\rm obs}-(\rm PSF\otimes \it 
        I_{\rm S,tot})_i |\rm ^2}{\sigma^2_{\rm tot,\it i}},
        \label{extsourcechi2}
\end{equation}
where $N_{\rm p}$ is the number of data pixels in the cutout, $\sigma_{\rm tot,\it i}$ is the uncertainty of pixel $i$. The symbol $\otimes$ represents the convolution of the PSF with the predicted S\'{e}rsic intensity $I_{{\rm S,tot}}$.

\begin{figure*}
        \centering
        \includegraphics[scale=0.55]{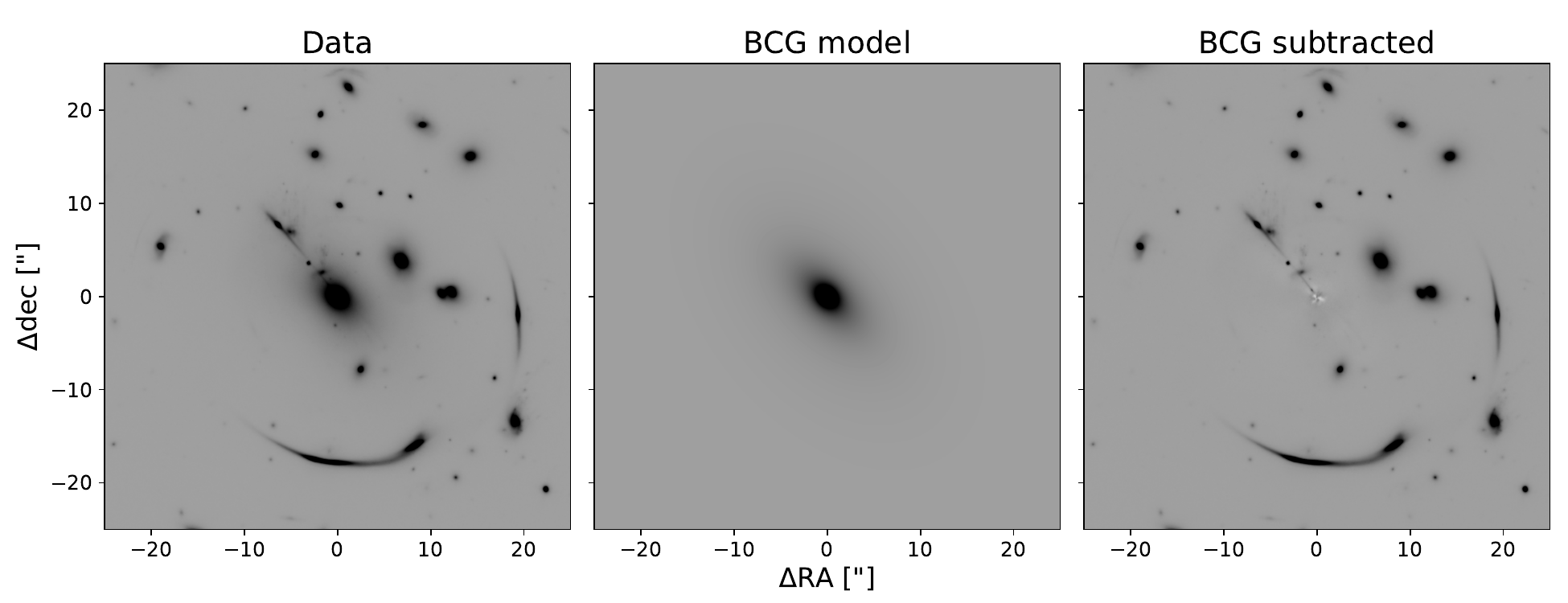}
        \caption{Results from modelling the BCG light with S\'{e}rsic profiles in the HST F160W filter. The left panel shows the observed image, the middle panel shows our BCG light model, and the right panel is the BCG subtracted image. The images cover the central $50\arcsec \times 50\arcsec$ around the BCG.}
        \label{fig:bcg_sub}
\end{figure*}

\subsection{BCG subtraction}
\label{sec:photometry_sersic:BCG}

\begin{figure*}
        \centering
        \includegraphics[scale=1, trim={0 2.8cm 0 2.5cm}, clip]{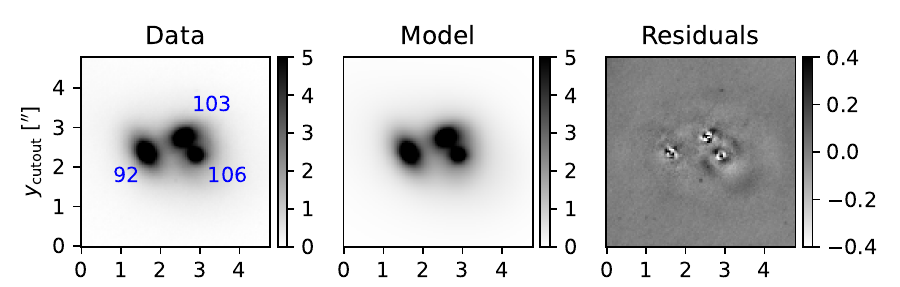}
        \includegraphics[scale=1.05, trim={0 2.9cm 0 3.2cm}, clip]{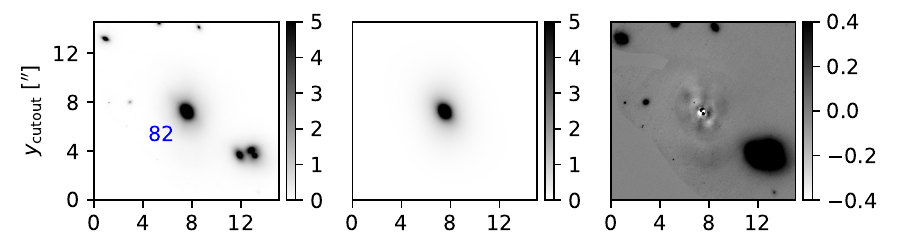}
        \includegraphics[scale=1.02, trim={0 2.3cm 0 3.2cm}, clip]{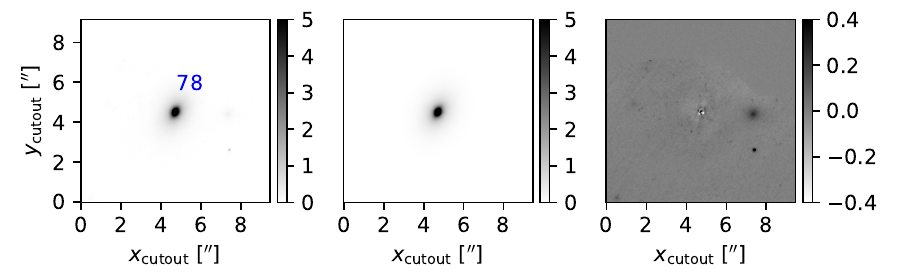}
        \caption{Examples for S\'{e}rsic fits to objects in the field in the JWST F200W filter.  From left to right: data image, model image, and residual image. The photometric IDs of the objects are shown in blue on the data image. Object 82 is a bright background galaxy at a redshift $z_{\rm bg} = 0.371$, and the other galaxies are cluster members close to the BCG.}
        \label{fig:sersic_res}
\end{figure*}

Brightest cluster galaxies are among the most massive and bright galaxies in the Universe. It is thus important to subtract off its light to avoid bias in the photometry of nearby cluster members and contamination of the giant arcs. In addition, clusters often exhibit intracluster light \citep[ICL;][]{Contini2021,Montes2022}, an extended and diffuse light envelope caused by unbound stars in the gravitational well of the DM halo. While BCGs can often be well fit with two S\'{e}rsic profiles to account for the inner and outer parts of the galaxy separately \citep{Gonzalez2005,Donzelli2011,Chu2021}, the disentanglement of BCG and ICL light is not straightforward \citep{Kluge2020,Kluge2021}. The ICL often has to be modelled as a separate component \citep[e.g.][]{JooJee2023}. We fitted the BCG light with three S\'{e}rsic profiles to also account for possible ICL contribution using the modelling software {\sc Glee} \citep[][]{Suyu.2010,Suyu+2012}. The light centroids are linked among the multiple S\'{e}rsic profiles. We found that we fitted the extended light distribution at the centre of the cluster slightly better (in terms of $\chi^2_{\rm SB}$ and by a visual check of the modelling residuals) with three S\'{e}rsic profiles instead of two.
In Fig.~\ref{fig:bcg_sub}, we show, as an example, the observed image, the BCG model, and the BCG subtracted image in the HST F160W filter, which is the filter we later used to obtain the cluster member scaling relations given the well-understood photometric calibrations of HST (Sect.~\ref{sec:mass_model:parametrizations}). At this wavelength, the BCG has a magnitude of $m_{\rm BCG,F160W}=15.30$ with a best-fit axis ratio of $q_{\rm BCG,F160W}=0.58$ and position angle $\phi_{\rm BCG,F160W}=53.2\degree$ (east of north), measured from its second brightness moments. The central region of the BCG exhibits prominent dust features, which leads to strong residuals in the central $\sim$1\arcsec, but overall the extended light is subtracted well. The subtraction also reveals more details of the radial arc and even a central image.

\subsection{S\'{e}rsic fitting of galaxies}
\label{sec:photometry_sersic:Sersic_fit}

We modelled the surface brightness of objects in the HST/JWST fields of the cluster with the software \texttt{Morphofit} \citep{Tortorelli2018,Tortorelli2023a,Tortorelli2023b}, which fits to galaxies in an automated and parallelized way. In \texttt{Morphofit}, the initial object detection and estimation of S\'{e}rsic parameter values is done with \texttt{SExtractor} \citep{Bertin1996} and the final S\'{e}rsic fit is conducted with the modelling software \texttt{Galfit} \citep{Peng2011}. The S\'{e}rsic parametrization is the same as in Eq.~(\ref{eq:sersic}). \texttt{Morphofit} will create cutouts of all objects and model them in parallel on multiple cores. We selected objects from the \texttt{SExtractor} catalogue with F160W magnitudes \texttt{MAG\_AUTO} $<23.5$, i.e., all objects with an automatically detected F160W magnitude fainter than \texttt{MAG\_AUTO}$=23.5$ are rejected to limit the number of objects in the mass model by including only the most massive ones. In a lens cluster at a similar redshift, \citet{Bergamini2023} identified a threshold magnitude of $m_{\rm F160W} \leq 21$; the inclusion of a fainter object (fainter than this threshold) in the mass model generally has no significant impact. Nevertheless we chose a uniform, inclusive approach and included all cluster members within the FOV that are above our magnitude threshold.

Initially, we modelled each galaxy with two S\'{e}rsic profiles with forced photometry (i.e., linking the centroid coordinates) among the multiple filters. Models with a single S\'{e}rsic component showed insufficient light subtraction, especially in the brighter central parts. Adding a second S\'{e}rsic component accounts for the bulge and disc decomposition that we observe in many of the galaxies in the field. For simplicity we decided to uniformly model all galaxies with two S\'{e}rsic profiles. After a visual quality check, we added a third S\'{e}rsic profile to the models that still show an excess of light in the residuals that could potentially impact the photometry. Other bright objects, which are sufficiently close, are fit simultaneously to avoid bias in the photometry. We fitted a total of 482 galaxies (in addition to the BCG) across all bands and estimated their total integrated AB magnitudes. All fitted objects are labelled uniquely with a photometric ID.
The magnitudes for the cluster members selected in Sect.~\ref{sec:selec:CM_selection} and the background galaxy (see Sect.~\ref{sec:selec:LOS}) are summarized in Table~\ref{tab:mag_cat}, which can be downloaded online\footnote{Link available upon publication.}.

In Fig.~\ref{fig:sersic_res}, we show as examples the modelled surface brightness and residuals for several objects close to the BCG. The three blended objects 92, 103, and 106 in the first row of the plot highlight the necessity of fitting S\'{e}rsic profiles to the galaxies instead of using automatic aperture photometry from \texttt{SExtractor} alone. We could deblend these objects and got reliable magnitudes for the individual galaxies, although there are some extended residual features, possibly due to tidal interactions of the galaxies. The second row shows the massive background galaxy at a redshift of $z_{\rm bg} = 0.371$, and the third row a bright cluster member just south of the BCG.

\section{Deflector galaxies and multiply lensed sources}
\label{sec:selec}

In our lens mass model of MACS J0138$-$2155, we included as deflectors spectroscopically confirmed and photometrically selected cluster members. In total, we found 84 objects at the cluster redshift of $z_{\rm d}=0.336$, including the BCG. We describe our cluster member selection in Sect.~\ref{sec:selec:CM_selection}. In addition to the mass at the cluster redshift, we included two line-of-sight (LOS) perturber galaxies, one in the foreground ($z_{\rm fg}=0.309$) and one in the background ($z_{\rm bg}=0.371$). We describe the LOS objects in Sect.~\ref{sec:selec:LOS}.
Our image-position constraints, summarized in Sect.~\ref{sec:selec:image_systems}, consist of 23 multiple images from eight image systems at four different source redshifts, selected from JWST imaging and MUSE spectroscopic data. 

\subsection{Cluster members}
\label{sec:selec:CM_selection}

Since MUSE covers only the central 1$\times$1~arcmin$^2$ region of MACS J0138$-$2155 and some cluster members are located outside this area, we used the photometric information of the galaxies described in Sect.~\ref{sec:photometry_sersic} to identify additional members. In particular, we used our spectroscopic catalogue to identify spectroscopic cluster members, and used both colour-magnitude and colour-colour diagrams to further select cluster members photometrically to complement the spectroscopic members.

\begin{figure*}
    \includegraphics[width=0.7\linewidth,  trim=3cm 0cm 3cm 0cm, clip, valign=c]{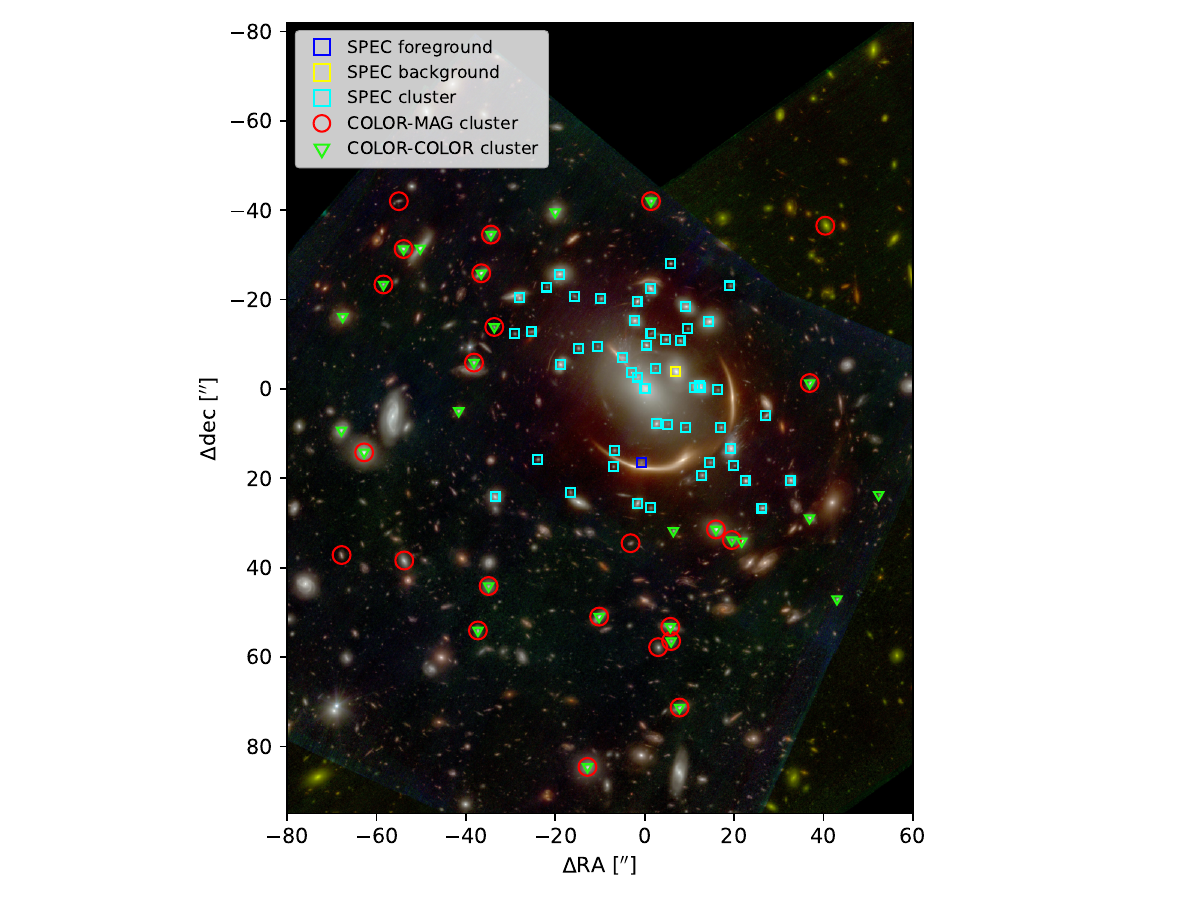}
    \hfill
    \begin{minipage}{\dimexpr 0.3\linewidth-\columnsep}
        \captionsetup{singlelinecheck=off, skip=0pt, justification=justified}
        \caption{Colour image of MACS J0138$-$2155 built from JWST NIRCam filters (red: F277W, F356W, F444W; green: F150W, F200W; blue: F115W). Spectroscopically confirmed objects are marked with squares, coloured in cyan for cluster members, and blue and yellow for the foreground and background objects, respectively. Objects selected from the colour-magnitude fit are marked as red circles, and green triangles are the colour-colour selections.}
        \label{fig:model_galaxies}
    \end{minipage}
\end{figure*}

\begin{figure}
        \centering
        \includegraphics[width = 1.0\columnwidth]{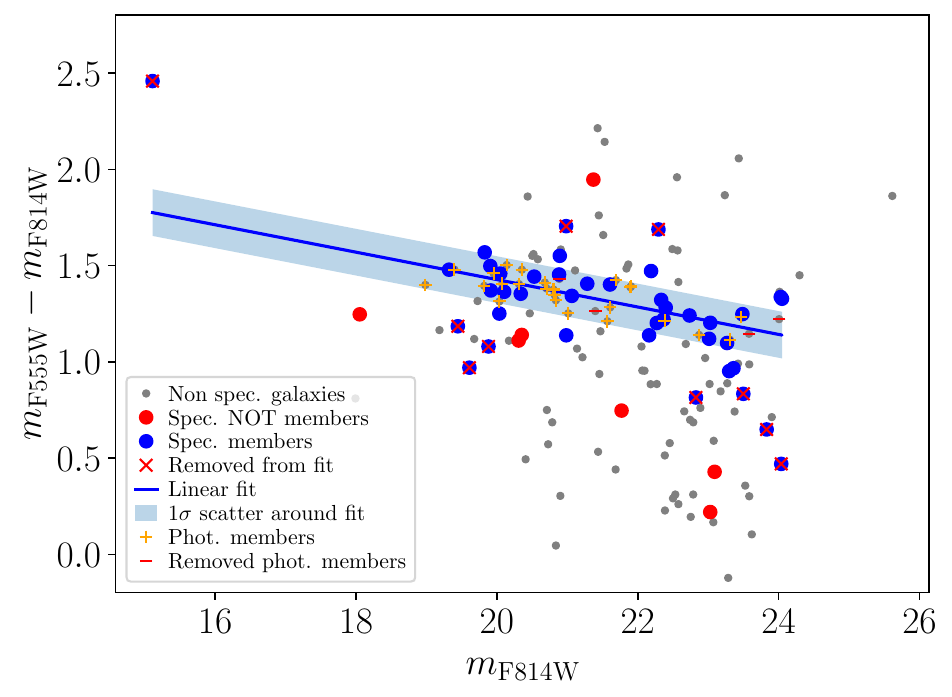}
        \caption{Colour-magnitude diagram in the field of MACS J0138$-$2155. Spectroscopically confirmed cluster members and non-members are indicated as blue and red points, respectively. Sources with no spectroscopic confirmation are shown in grey. The blue line and region indicate the fitted red sequence and the 1$\sigma$ scatter. Red crosses (X) mark cluster members that do not follow the red sequence and yellow plus signs indicate the photometrically selected members that are included in our lens model. Objects inside the red sequence with a red minus sign are not included as photometric cluster members due to the lack of, or unreliable, F160W photometry.}
        \label{fig:color_mag_members}
\end{figure}

We used the sample of 50 spectroscopically confirmed cluster galaxies in the redshift range $z=0.324-0.349$ identified by \citet{Granata+24} as the basis for our catalogue of members. These galaxies are marked by cyan squares on the JWST image in Fig.~\ref{fig:model_galaxies}. After excluding the three jellyfish galaxies\footnote{The spectroscopic IDs of these three jellyfish galaxies are 100001, 100002, and 100003 in the spectroscopic catalogue presented by \citet{Granata+24}. }, we used the remaining spectroscopic objects to define the red sequence in the colour-magnitude diagram shown in Fig.~\ref{fig:color_mag_members}. We used the F555W $-$ F814W colour from HST, given that these two filters roughly bracket the rest-frame 4000\,$\rm{\AA}$ break of the cluster members, providing a tighter red sequence that is helpful for selecting additional members photometrically.
The spectroscopically confirmed cluster members are shown as blue points, whereas spectroscopically confirmed non-members are red.  Galaxies with photometric measurements from Sect.~\ref{sec:photometry_sersic} and without spectroscopic redshifts are in grey.  

We fitted a linear relation to the spectroscopic cluster members (blue points) through a least-squares fit and obtained the 1$\sigma$ uncertainty on this linear relation (the blue line and the shaded blue region).  We performed a $\sigma$-clipping (at 2$\sigma$) to exclude outliers in spectroscopic members, which are marked by red crosses\footnote{We use 2$\sigma$ as a threshold since a 3$\sigma$ threshold would include spectroscopic non-members in the selection, which we would like to avoid.}. The BCG, which is more than three magnitudes brighter than the other cluster members in this cluster, was clipped in this process. We then repeated the fitting and $\sigma$-clipping until no additional spectroscopic cluster members are detected/excluded as outliers, i.e., all remaining spectroscopic cluster members lie within 2$\sigma$ of the linear fit (within twice the thickness of the light-shaded blue 1$\sigma$ region). 

All the galaxies without spectroscopy, but with colours and magnitudes lying within 1$\sigma$ of the red sequence (in shaded blue), were selected as photometric cluster members and are marked by yellow plus signs in Fig.~\ref{fig:color_mag_members}. In total, 28 galaxies were selected as cluster members with the colour-magnitude method. We however excluded four galaxies that are marked by red minus signs in the red sequence since we cannot readily use their photometry in the scaling relations of elliptical galaxies -- one of them is a spiral galaxy, one is outside the F160W FOV without F160W photometry, and two of them have unreliable photometry from the elliptical S\'{e}rsic fit (one is located at the edge of the WFC3 chip and one has non-elliptical morphology). The resulting 24 colour-magnitude selected cluster members are marked by red circles on the JWST image shown in Fig.~\ref{fig:model_galaxies}.

\begin{figure*}
        \centering
        \includegraphics[width = 1\columnwidth]{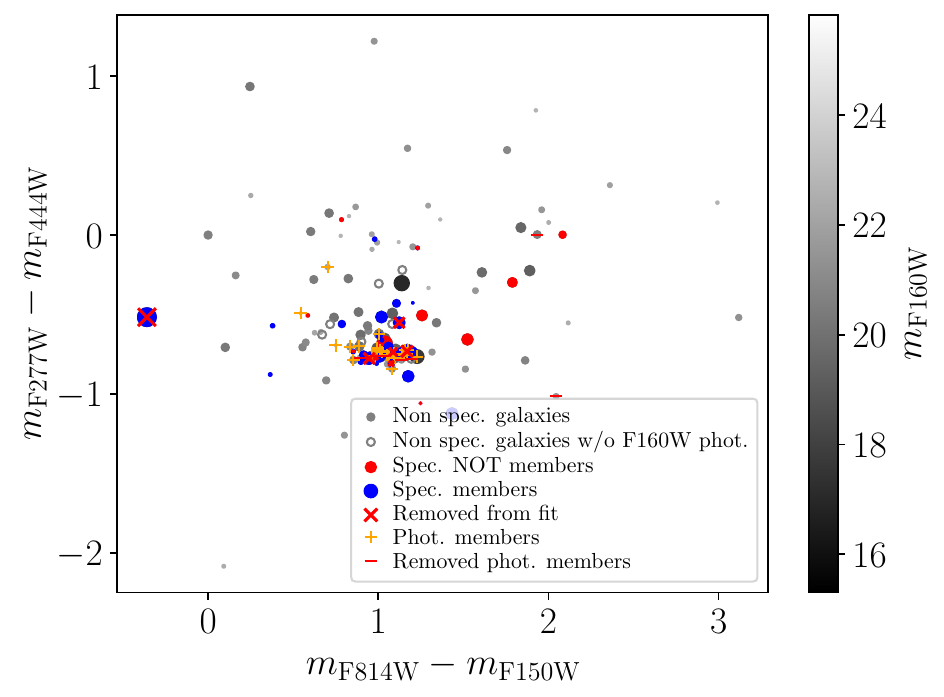}
        \includegraphics[width = 1\columnwidth]{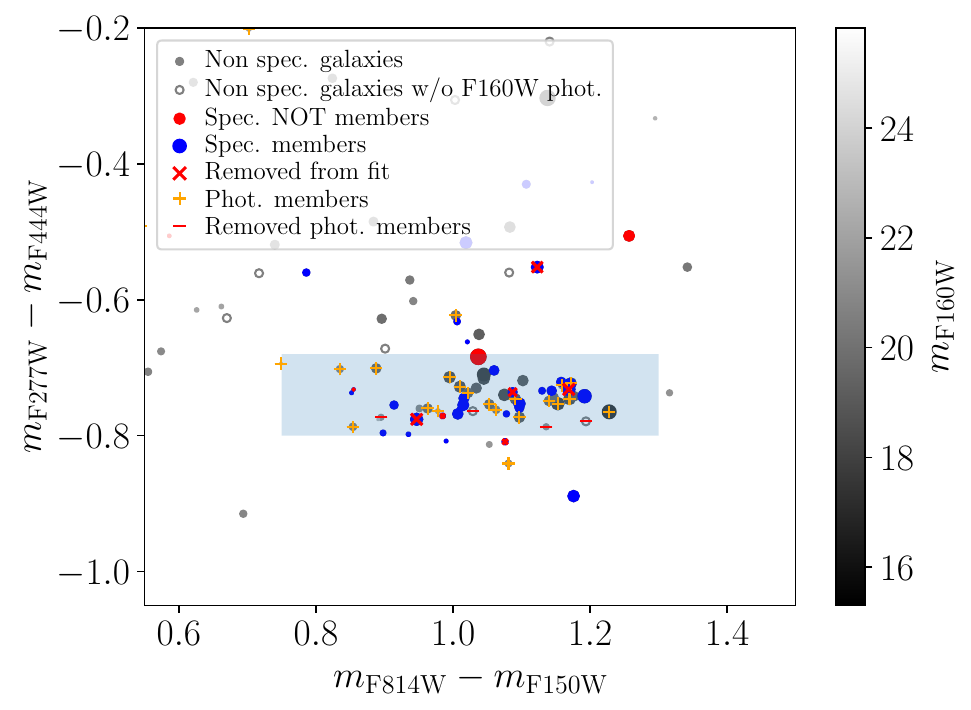}
        \caption{Colour-colour diagram in the field of MACS J0138$-$2155, where the right panel is a zoom-in of the left panel. Spectroscopically confirmed cluster members and non-members are indicated in blue and red, respectively. Sources with no spectroscopic confirmation are shown in grey, with size and shade related to the galaxy brightness as shown in the colour bar. Red crosses (X) mark spectroscopic cluster members that do not follow the red sequence and yellow plus signs indicate the colour-magnitude selected members (from Fig.~\ref{fig:color_mag_members}) that are included in our lens model. Photometric galaxies inside the blue shaded region in the right-hand panel are selected as cluster members via our colour-colour selection.}
        \label{fig:color_color_members}
\end{figure*}

We supplemented the colour-magnitude selection of cluster members by using additional colour-colour selection.  Following \citet{Frye2024}, we used the F277W$-$F444W and F814W$-$F150W colours, where we substituted the JWST F090W filter in \citet{Frye2024} with the HST F814W filter due to their similar wavelengths and the lack of JWST F090W observations for SN Encore. Figure~\ref{fig:color_color_members} shows the colour-colour diagram on two different scales, where the right-hand panel is a zoom-in of the left-hand panel. The symbols are the same as in Fig.~\ref{fig:color_mag_members}, where red points are spectroscopic non-members, blue points are spectroscopic members, grey circles are objects without spectroscopic redshifts, red crosses are spectroscopic members that are not within the red-sequence region, yellow plus signs are photometric members (grey) that lie within the red sequence and were selected as cluster member, and red minus signs are objects not included as cluster members due to unreliable/missing F160W photometry. The sizes/transparency of the filled grey circles are related to the brightness of the galaxies, where brighter galaxies in F160W are represented by bigger and darker dots (as shown by the colour bar).  Open grey circles are galaxies without F160W photometry.  In the right-hand panel, we define a region with $-0.80 < {\rm F277W - F444W} < -0.68$ and $0.75 < {\rm F814W - F150W} < 1.3$ to include photometric galaxies (filled grey circles) as cluster members, some of which are on the brighter side.  

We tuned this colour-colour selection region (light blue shaded region) to include the brighter photometric members while excluding the spectroscopic non-members.  In total, we selected 32 photometric galaxies via the colour-colour diagram, of which 14 are additional cluster members that were not previously selected through the colour-magnitude diagram. However, two of the 14 galaxies do not have HST F160W photometry as they are outside the F160W field of view, and two other galaxies are located very close to the edge of the FOV.  We excluded these four galaxies (marked by red minus signs in Fig.~\ref{fig:color_color_members}) from the model since we cannot assign mass to them via scaling relations without the F160W photometry. Their effect on the mass model is likely to be minimal given their faintness compared to other cluster members and their large distance to the cluster core. The 28 colour-colour selected galaxies are marked by triangles in Fig.~\ref{fig:model_galaxies}.

As visible in Fig.~\ref{fig:model_galaxies}, our spectroscopic, colour-magnitude, and colour-colour selection of cluster members include significantly massive early-type galaxies that are crucial for incorporating into the lens mass model, especially those galaxies near the cluster centre where we have lensed multiple image systems. The union of the colour-magnitude and colour-colour selections ensures a higher completeness in the cluster members, especially for including a few of the massive cluster member that are missed in one of the selections. The objects that are close to but outside of the borders of the selection regions (either in colour-magnitude in Fig.~\ref{fig:color_mag_members} or colour-colour in Fig.~\ref{fig:color_color_members}) are either faint or substantially distant from the cluster centre, and their lensing effects are expected to be negligible. In total, we included 50 spectroscopically selected, and 34 photometrically selected cluster members in our model.

\subsection{LOS objects}
\label{sec:selec:LOS}

We checked for potentially significant line-of-sight objects for cluster mass modelling, i.e. galaxies near arcs or that appear to be massive.  We spectroscopically confirmed a small foreground galaxy close to the giant arc at a redshift of $z_{\rm fg}=0.309$ (RA, dec = 24.5159632\degree, $-$21.9300802\degree), and a massive background galaxy at $z_{\rm bg}=0.371$ (RA, dec = 24.5136457\degree, $-$21.9244117\degree), about 8\arcsec\ northwest of the BCG. We included both into our mass model given their proximity to the giant arcs or the high mass of the background galaxy (based on its stellar velocity dispersion measurement). The final selection of galaxies to be included in the model is shown in Fig.~\ref{fig:model_galaxies}. This includes cluster members, as well as LOS objects.

\subsection{Strongly lensed background sources as constraints}
\label{sec:selec:image_systems}

We securely identified eight multiple image systems from four independent background galaxies, with a total of 23 images in the redshift range of $0.767\leq z \leq3.420$, which are summarized in Table~\ref{tab:img_systems}. These multiple image systems were voted on and agreed upon by eight groups of researchers\footnote{Two of the groups using the same lens-modelling software subsequently merged during the modelling stage, which resulted in seven independent lens mass models.}, as described in Suyu et al. (in prep.). Each multiple image system was labelled numerically, and the individual lensed images of each system were labelled alphabetically, in decreasing order of brightness.  The multiple images of SN Encore (JWST) and SN Requiem (HST) are Systems 1 and 2, respectively. System 3 is associated with the corresponding emissions from the centre of the SN host galaxy at $z=1.949$. The positions were identified from the JWST images. Systems 4.1, 4.2, and 4.3 correspond to [OII] emissions at $z_{\rm s,4}=0.767$ identified in the MUSE data, where different clumps of emission are lensed into multiple image systems; since these emissions likely come from the same galaxy, we labelled them with the same starting number 4. The image positions of Systems 4.1 and 4.3 are based on MUSE data, whereas System 4.2 is based on JWST images. Additionally, two Ly-$\alpha$ emission regions were found in MUSE at a redshift of $z_{\rm s,5}= 3.152$ and $z_{\rm s,6}= 3.420$, which we labelled as Systems 5 and 6, respectively; these two Ly-$\alpha$ sources do not have clear detections in the JWST imaging. 
We assigned elliptical errors on their positions based on the spatial extent of the images in the data (in either JWST or MUSE data, depending on the data in which the image positions were detected/measured).

\begin{figure*}
    \includegraphics[width=0.7\linewidth, valign=c]{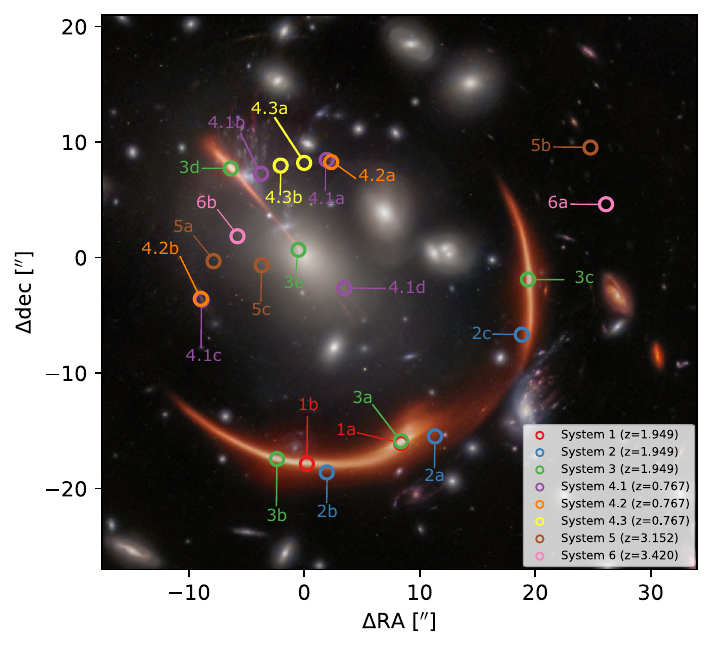}
    \hfill
    \begin{minipage}{\dimexpr 0.3\linewidth-\columnsep}
        \captionsetup{singlelinecheck=off, skip=0pt, justification=justified}
        \caption{Positions of the eight multiple image systems (the `gold' systems) used to constrain our mass model. We show the positions in arcseconds relative to the BCG at (RA, dec) = (24.5157084\degree, $-$21.9254717\degree).}
        \label{fig:imgpos}
    \end{minipage}
\end{figure*}

In addition to these eight multiple image systems that were designated as ``gold" systems, there is one more lensed image system that was designated as ``silver" (where we refer to Suyu et al. (in prep.) for the criteria of gold and silver systems).  The silver system corresponds to the spectroscopic ID 1755 \citep{Granata+24}, with a ``likely" instead of a ``secure" redshift measurement coming from two blended background sources. Furthermore, the source is primarily lensed by a single cluster member rather than the full cluster, so the lensing configuration is mainly informative on the total mass distribution of the specific cluster member rather than the global cluster mass distribution.  This specific cluster member is $\gtrsim$17\arcsec\ away from all the SN images and therefore has negligible impact on the SN image properties (of both SN Encore and SN Requiem).  Therefore, we did not include this silver system into our mass modelling. Figure~\ref{fig:imgpos} shows an overview of the eight multiple image systems and their redshifts.

\begin{table*}
  \caption[Observed multiple image positions and errors.]{Observed multiple image positions and positional uncertainties.}
  \label{tab:imagepos}
  \begin{center}
  \begin{tabular}{lccc}
    \hline
    \\
     & $x\ ['']$ & $y\ ['']$ & elliptical positional uncertainty \\
    \\
    \hline
    System 1 ($z_{\rm s,1}= 1.949$, SN Encore) & & &\\
    \hline
    \addlinespace[2pt]
    Image 1a & \phantom{$-$}8.396  &    $-$16.065  & 0.04, 0.04, 0\\
    Image 1b & \phantom{$-$}0.233   &    $-$17.864 &  0.04, 0.04, 0\\
    \hline
    System 2 ($z_{\rm s,2}= 1.949$, SN Requiem) & & &\\
    \hline
    Image 2a & \phantom{$-$}11.329  &    $-$15.482  & 0.08, 0.08, 0\\
    Image 2b & \phantom{$-$}1.955   &    $-$18.604 &  0.08, 0.08, 0\\
    Image 2c & \phantom{$-$}18.843  &     $-$6.677 &  0.08, 0.08, 0\\
        \hline
    System 3 ($z_{\rm s,3}= 1.949$) & & &\\
    \hline
    Image 3a& \phantom{$-$}8.379  &  $-$15.899 & 0.08, 0.04, 135\\
    Image 3b & $-2.355$  &  $-$17.445 & 0.16, 0.04, 80\\
    Image 3c& \phantom{$-$}19.378  &   $-$1.900 & 0.04, 0.08, 90\\ 
    Image 3d& $-6.343$   &   \phantom{$-$}7.713 & 0.06, 0.12, 135\\
    Image 3e& $-0.500$   &   \phantom{$-$}0.679 &  0.06, 0.12, 118\\
     \hline
    System 4.1 ($z_{\rm s,4.1}= 0.767$) & & &\\
    \hline
    Image 4.1a & \phantom{$-$}1.959  &    \phantom{$-$}8.454  & 0.6, 0.4, 75.3\\
    Image 4.1b & $-$3.720   &    \phantom{$-$}7.260 &  0.6, 0.4, 154.6\\
    Image 4.1c & $-$8.870  &     $-$3.653 &  0.6, 0.4, 45\\
    Image 4.1d & \phantom{$-$}3.488  &     $-$2.648 & 0.6, 0.6, 0\\
        \hline
        System 4.2 ($z_{\rm s,4.2}= 0.767$) & & &\\
    \hline
    Image 4.2a & \phantom{$-$}2.340  &    \phantom{$-$}8.266  & 0.2, 0.08, 62.1\\
    Image 4.2b & $-$8.949   &    $-$3.570 & 0.2, 0.08, 26.7\\
            \hline
                    System 4.3 ($z_{\rm s,4.3}= 0.767$) & & &\\
    \hline
    Image 4.3a & $-$0.003  &    \phantom{$-$}8.193  & 0.6, 0.3, 84.5\\
    Image 4.3b & $-$2.038   &    \phantom{$-$}7.959 & 0.6, 0.3, 117.8\\
            \hline
        System 5 ($z_{\rm s,5}= 3.152$) & & &\\
    \hline
    Image 5a & $-$7.844  &    $-$0.313  & 0.5, 0.3, 106.8\\
    Image 5b & \phantom{$-$}24.787   &    \phantom{$-$}9.526 &  0.4, 0.4, 0\\
        Image 5c & $-$3.677   &    $-$0.662 &  0.5, 0.3, 51.4\\
        \hline
        System 6 ($z_{\rm s,6}= 3.420$) & & &\\
    \hline
    Image 6a & \phantom{$-$}26.135  &    \phantom{$-$}4.617  & 0.4, 0.4, 0\\
    Image 6b & $-$5.773   &    \phantom{$-$}1.846 &  0.4, 0.4, 0\\
    \hline

  \end{tabular}
  \end{center}
  \caption*{Notes: In this table, we provide the $x$ and $y$ coordinates
    with respect to the BCG (RA=24.51570318\degree, dec=$-$21.92547911\degree), and the relative elliptical errors in the format major axis $['']$, minor axis $['']$, position angle [\degree], with position angle measured counterclockwise from the positive $y$-axis (east of north).}
  \label{tab:img_systems}
  \end{table*}

\section{Lens mass models}
\label{sec:mass_model}
Having selected the cluster members, identified important LOS objects, and determined the multiple image systems, we constructed our cluster lens mass model. We describe the parametrization of the different mass components in Sect.~\ref{sec:mass_model:parametrizations}, and present seven different {\sc Glee} mass models in Sect.~\ref{sec:mass_model:models} in order to quantify systematic uncertainties due to mass modelling assumptions.  We then describe the methods of modelling MACS J0138$-$2155 with the multiple image systems as constraints in Sect.~\ref{sec:mass_model:img_pos_modeling}. The general modelling methodology is very similar to that developed by \citet{Grillo2015} and \citet{Grillo2016} for two lens galaxy clusters of the Hubble Frontier Fields \citep{Lotz2017}.

\subsection{Mass parametrizations}
\label{sec:mass_model:parametrizations}

\subsubsection{Galaxy mass distributions}
\label{sec:mass_model:parametrizations:galaxies}

We modelled the cluster members and LOS galaxies with a truncated dual pseudo-isothermal elliptical mass distribution \citep[dPIE;][]{Eliasdottir07, SuyuHalkola2010} with vanishing core radius. Their dimensionless projected surface mass density, i.e., convergence, is
\begin{equation}
\label{eq:M0138kappadPIE}
\scalebox{1.1}{$ \kappa_{\rm dPIE}(x,y) \Big|_{\rm z_s=\infty}= \frac{\theta_{\rm E,\infty}}{2} \left( \frac{1}{R_{\rm em}} - \frac{1}{\sqrt{R_{\rm em}^2+r_{\rm t}^2}}\right) ,$}
\end{equation}
where the 2D elliptical mass radius with respect to their centre is
\begin{equation}
\label{eq:M0138rem}
\scalebox{1.2}{$ R_{\rm em}(x,y)=\sqrt{\frac{x^2}{(1+e)^2}+\frac{y^2}{(1-e)^2}}, $}
\end{equation}
and the ellipticity, $e,$ is related to the axis ratio, $q,$ by
\begin{equation}
\label{eq:M0138e}
\scalebox{1.3}{$e=\frac{1-q}{1+q}$}.
 \end{equation}
The parameter $\theta_{\rm E,\infty}$ is the Einstein radius for a source at redshift $z_{\rm s}=\infty$, and $r_{\rm t}$ is the truncation radius. The mass distribution is then rotated by its position angle $\phi$. We fixed the coordinates of each object on the image plane to the observed positions from the HST F160W image (from the S\'{e}rsic fit described in Sect.~\ref{sec:photometry_sersic}). For simplicity, we
assumed all the cluster members and LOS perturbers to be spherical, with the exception of the BCG for which we considered different scenarios, a spherical BCG and an elliptical BCG, given the substantial mass of the BCG.

To reduce the number of free parameters in our model (such that the number of free parameters does not exceed the number of observables) and increase the computational efficiency of our modelling, we adopted scaling relations for the Einstein and truncation radii of the cluster members \footnote{The scaling relation is only applied to cluster members and not to LOS galaxies.}, i.e., we scaled them with respect to those of a reference galaxy. The choice of the reference galaxy does not impact the result of the modelling. In our case, we chose to scale the values of $\theta_{\rm E,\infty}$ and $r_{\rm t}$ with power laws with respect to the luminosity of the member using object ID$_{\rm phot}=$115 as a reference ($m_{\rm F160W,\ ID115}=17.99$), so that $\theta_{\rm E,\infty}$ and $r_{\rm t}$ for the $i$-th member are
\begin{equation}\label{scaling1}
       \theta_{{\rm E,\infty},i} = \theta_{\rm{ E},\infty}^\mathrm{ref} \left(\frac{L_i}{L_\mathrm{ID 115}}\right)^{2 \alpha}, \\
\end{equation}
\begin{equation}\label{scaling2}
       r_{{\rm t},i}= r^\mathrm{ref}_{\rm t} \left(\frac{L_i}{L_\mathrm{ID 115}}\right)^{\beta}, \\
\end{equation}
where $L_i$ and $L_\mathrm{ID 115}$ are measured in the HST band F160W, and $\alpha$ and $\beta$ are power-law indices that we determine through properties of cluster members, particularly velocity-dispersion measurements. 

Cluster-scale strong lensing is sensitive to the total mass of the member galaxies, but unless galaxy-scale lensing systems are observed \citep{Granata23,Galan24}, the reconstruction of their mass structure is affected by a degeneracy between the values of the Einstein and truncation radii of the members, which affects our predictions on their compactness. Several recent works \citep{Bergamini19,Bergamini21,Granata22} have shown that this degeneracy can be broken with the introduction of independent kinematic priors on Eq. (\ref{scaling1}). 

The value of $\theta_{\rm E,\infty}$ is linked to the velocity dispersion parameter of the dPIE profile  \citep{Eliasdottir07} by
\begin{equation}\label{sigmatore}
    \theta_{\rm E,\infty}=4 \pi \left(\frac{\sigma_\mathrm{dPIE}}{c}\right)^2.
\end{equation}
For a vanishing core radius, $\sigma_\mathrm{dPIE}$ is well approximated by the value of the line-of-sight stellar velocity dispersion measured within a small aperture \citep[smaller than the truncation radius, ][]{Bergamini19}. We know that a power-law relation \citep[also known as the Faber-Jackson law, ][]{Faber76} holds between the measured stellar velocity dispersion of elliptical galaxies and their total luminosity.
\citet{Granata+24} provided us with a sample of 13 passive cluster galaxies in MACS J0138$-$2155 with reliable velocity-dispersion measurements within their effective radii. We can thus calibrate the Faber-Jackson law for the cluster members
\begin{equation}\label{fj}
       \sigma_{{\rm v},i} = \sigma_{\rm v}^\mathrm{ref} \left(\frac{L_i}{L_\mathrm{ID 115}}\right)^{\alpha},\end{equation}
where $\alpha$ was defined in Eq. (\ref{scaling1}). From Eq. (\ref{sigmatore}), we derive $\theta_{\rm E,\infty}^\mathrm{ref} = 4 \pi \left(\frac{\sigma_v^\mathrm{ref}}{c}\right)^2$. \citet{Granata+24} fit Eq. (\ref{fj}) adopting a Bayesian technique analogous to \citet{Bergamini19}. They find $\alpha=0.25^{+0.05}_{-0.05}$ and $\sigma_{\rm v}^\mathrm{ref}=206^{+14}_{-13} \, \mathrm{km \, s^{-1}}$, which is compatible with the stellar velocity dispersion value for the reference member ID 115 ($211.1 \pm 3.1 \, \mathrm{km \, s^{-1}}$); the scatter is $\Delta \sigma_v = 25 ^{+6}_{-4} \, \mathrm{km \, s^{-1}}$ for the Faber-Jackson relation, which we used to estimate the mean and standard deviation of the Gaussian prior on $\theta_{\rm E,\infty}^\mathrm{ref}$. For the second scaling relation in Eq. (\ref{scaling2}), we chose $\beta=0.7$, so that $M_{\mathrm{tot},i}/L_{i} \propto L_{i}^{0.2}$, consistent with the observed Fundamental Plane relation. 

By varying only two parameters, i.e. the Einstein and truncation radii of our reference galaxy, we determined the total mass distribution of all cluster members that follow the scaling relations.  In practice, we excluded the BCG and object ID$_{\rm phot}=$116 from the scaling relations and used Gaussian priors on their Einstein radii from their measured velocity dispersions. The BCG is too bright to follow the scaling relation; the galaxy with ID$_{\rm phot}=$116 is close to the critical curve of image systems 4.1, 4.2 and 4.3 and the predicted image multiplicity and positions are very sensitive to the mass distribution of this cluster member. While we assumed that the BCG is spherical in most of our mass models, we also considered an elliptical BCG in one of our mass models.  Furthermore, the jellyfish galaxies are free to vary outside the scaling relation as well, with their truncation radii fixed to 5\arcsec\ and a uniform prior on their Einstein radii 
(see Tables~\ref{tab:mass_models} and \ref{tab:encore_final_param} for details on the assigned priors on the mass components).

For the two LOS galaxies (see Sect.~\ref{sec:selec:LOS}), we fixed their centroids to the observed light distributions.  The foreground galaxy does not have reliable photometry, and given its proximity to the southern giant arc, we allowed both its Einstein and truncation radii to vary.  The background galaxy has a measured velocity dispersion, which we used to impose a prior on its Einstein radius.

\subsubsection{Cluster DM halo mass distributions}
\label{sec:mass_model:parametrizations:clusterhalo}

We included two cluster DM (dark matter) haloes in our model, since we found that a single cluster DM halo cannot reproduce the positions of the eight multiple image systems well.  We found that the primary DM halo is more centrally concentrated and massive, whereas the secondary DM halo is extended and cored.  The multiple image systems require that the centroid of the secondary DM halo to be offset from that of the primary DM halo. As will be seen in Sect.~\ref{sec:results_fixed_cosmo}, the secondary halo is a perturbation of the primary halo to allow for non-elliptical halo mass distribution.  Hereafter, we refer to the secondary halo as a perturbative halo that alters the primary halo structure, as opposed to a separate and physically present DM halo. For the primary DM halo, we considered three possible profiles: (1) pseudoisothermal elliptical mass distribution \citep[PIEMD;][]{Kassiola.1993}, (2) softened power-law elliptical mass distribution \citep[SPEMD;][]{Barkana+99}, and (3) elliptical Navarro Frenk \& White profile \citep[NFW;][]{Navarro+97,Oguri21}.  For the perturbative DM halo, we modeled it with a PIEMD.

The convergence for a PIEMD is
\begin{equation}
\label{eq:kappaPIEMD}
\kappa_{\rm PIEMD}(x,y)\Big|_{\rm z_s=\infty}= \frac{\theta_{\rm E,\infty}}{2 \sqrt{R_{\rm em}^2 +r_{\rm c}^2}},  
\end{equation}
where $R_{\rm em}$ is the elliptical mass radius (Eq.~\ref{eq:M0138rem}), $e$ is the ellipticity (Eq.~\ref{eq:M0138e}), $\theta_{\rm E, \infty}$ is the lens Einstein radius for a source at redshift infinity and $r_{\rm c}$ is the core radius. The mass distribution is then rotated by its position angle $\phi$, and centred at ($x_{\rm DM}$, $y_{\rm DM}$). 
All six parameters are free to vary within flat priors.

The convergence of a SPEMD is
\begin{equation}
\label{eq:kappaSPEMD}
\kappa_{\rm SPEMD}(x,y)= E \left(u^2+s^2\right)^{-\gamma},
\end{equation}
where $u^2=x^2+y^2/q^2$, $q$ is the axis ratio, $s = 2 r_{\rm c}/(1+q)$, $r_{\rm c}$ is the core radius, $\gamma$ is the radial power-law slope ($\gamma=0.5$ corresponds to an isothermal profile), and $E$ is the strength/amplitude of the lens.  The mass distribution is centred at ($x_{\rm DM}$, $y_{\rm DM}$) and has a position angle $\phi$.  The SPEMD profile therefore has seven free parameters, and we adopt flat priors for these parameters.

The three-dimensional mass density distribution of the NFW profile is
\begin{equation}
\label{eq:kappaNFW}
\rho_{\text{NFW}}(r) = \frac{\rho_{\rm s}}{(r/r_{\rm s})(1 + r/r_{\rm s})^2},
\end{equation}
where $r$ is the radial coordinate, $\rho_{\rm s}$ is the characteristic density, and $r_{\rm s}$ is the scale radius.  This can be projected to obtain the convergence \citep[e.g.][]{GolseKneib02}. The elliptical NFW convergence has six free parameters with flat priors: centroid ($x_{\rm DM}$, $y_{\rm DM}$), axis ratio ($q$), position angle ($\phi$), normalization ($\kappa_0$), and a scale radius ($r_{\rm s}$). We followed \citet{Oguri21} for calculating the lensing deflection angles and potentials of elliptical NFW convergence, and used the implementation from Wang et al. (in prep.).

\subsubsection{Mass sheet}
\label{sec:mass_model:parametrizations:sheet}

Since the mass-sheet degeneracy \citep[MST;][]{Falco+85, Gorenstein+88} is one of the dominant sources of uncertainty in time-delay cosmography, we also considered adding a constant sheet of mass parameterized by $\kappa_0$ in our mass model.  The value of $\kappa_0$ is not constrained by a single background source, since any value of $\kappa_0$ can produce a new mass distribution that fits the image positions equally well.  This new mass distribution consists of adding $\kappa_0$ at the cluster redshift while reducing the mass of the other components such as dark matter haloes and galaxies, and produces the same predictions of lensed image positions and relative magnifications of a single source \citep[e.g.][]{Schneider19}. In our case with four sources at different redshifts, $\kappa_0$ can no longer take an arbitrary value and still fit to the observed image positions.  The model with the mass sheet therefore allows us to quantify model uncertainties due to the mass-sheet degeneracy (as first shown in \citealt{Grillo2020}).

\subsection{Mass models}
\label{sec:mass_model:models}

We considered seven different GLEE mass models of the galaxy cluster in order to quantify uncertainties due to mass modelling assumptions.  We summarize the seven models in Table \ref{tab:mass_models}, including the number of variable mass model parameters and also indicating the priors that are imposed on some of the parameters based on the kinematic measurements of \citet{Granata+24}. 

The first four models explore the variety of plausible cluster DM halo mass distributions given its dominant role in reproducing the observed image positions.  In particular, the primary DM halo is described by a cored-isothermal profile (i.e., PIEMD) in the \texttt{iso\_halo} model, a power-law profile (i.e., SPEMD) in the \texttt{PL\_halo} model, and a NFW profile in the \texttt{NFW\_halo} model. The \texttt{iso\_halo+sheet} model is parameterized in the same way as the \texttt{iso\_halo} model with the addition of a mass sheet with the amplitude $\kappa_0$ as an extra free parameter.  The next model, \texttt{iso\_halo+ell\_BCG}, is similar to the \texttt{iso\_halo} model except that the BCG is described by an elliptical isothermal profile instead of a circular one.  This assesses the impact of assuming the BCG to be circular.  All of these five models have deflectors at different redshifts, given the presence of the foreground and background LOS galaxies, and employ multi-plane lens modelling \citep[e.g.][]{Wong2017, Chirivi+18, Wang2022, Acebron+24, Schuldt+24}.  These models also use the elliptical positional uncertainties listed in Table~\ref{tab:imagepos}.

The next \texttt{single\_plane} model is similar to the \texttt{iso\_halo} model but has the foreground galaxy at $z_{\rm fg}=0.309$ (near the southern giant arc) and the bright background galaxy at $z_{\rm bg}=0.371$ placed at the cluster redshift of $z_{\rm d}=0.336$, which reduces the model to a single deflector plane.  This helps to assess whether it is important to use multi-plane lensing for this cluster.

The last `\texttt{baseline}' model is considered in order to facilitate direct comparison of modelling results from the various independent modelling teams, which are presented in Suyu et al. (in prep.).  Since the teams use different modelling software with varying capabilities, not all software have multi-plane lensing or the usage of elliptical positional uncertainties (instead of circular uncertainties).  Therefore, the teams agreed to build the so-called \texttt{baseline} model with all the deflectors assumed to be at the cluster redshift and all the positional uncertainties to be circular, which is a setup that all the teams can perform.  For each multiple image position, the size of the circular uncertainty ($\sigma^{\rm circ}$) is the geometric mean of the semi-major ($\sigma^{\rm major}$) and semi-minor ($\sigma^{\rm minor}$) axes of the elliptical uncertainty tabulated in Table~\ref{tab:imagepos}, i.e., $\sigma^{\rm circ} = \sqrt{\sigma^{\rm major} \sigma^{\rm minor}}$. Such a baseline model will enable a more direct comparison of the modelling software and methodology.  Each team can then build their `ultimate' models, relaxing assumptions from the baseline model that they would use for cosmographic analysis.  

Having presented the various mass model parameterizations that we have considered, we describe next our procedure to constrain the model parameter values based on the image positions of the multiple image systems.

\subsection{Mass modelling with lensed image positions}
\label{sec:mass_model:img_pos_modeling}
Our lens mass modelling of MACS J0138$-$2155 was performed with \GLEE\ \citep{SuyuHalkola2010,Suyu+2012}. We first optimized the model with the multiple image positions described in Sect.~\ref{sec:selec:image_systems} by minimizing the angular separation between the observed, $\vec{\theta}^{\rm obs}$, and model-predicted, $\vec{\theta}^{\rm pred}$, image positions on the image plane.
The model parameters $\boldsymbol{\eta}$ (vector of length $N_{\rm par}$, the number of lens parameters) are varied to minimize the $\chi^2$ function
\begin{equation}
        \chi^2_{\rm im} = \sum_{j=1}^{N_{\rm sys}}\sum_{i=1}^{N_{\rm im}^{j}}  \left (
        \Delta \vec{\theta}_{ij} \right )^{\rm T} {\mathsf{R}}{\mathsf{C_{im}}}^{-1}{\mathsf{R}}^{\rm T} \left (
        \Delta \vec{\theta}_{ij} \right ),
        \label{eq:chi2im_ell}
\end{equation}
where $N_{\rm sys}$ is the number of multiple-image systems, $N_{\rm im}^j$ is the number of multiple images in image system $j$, and the image positional offset is 
\begin{equation}
        \Delta \vec{\theta}_{ij} = \vec{\theta}_{ij}^{\rm obs}-\vec{\theta}_{ij}^{\rm pred}(\boldsymbol{\eta}, \vec{\beta}_{j}^{\rm mod}).
\end{equation}
The two-dimensional vectors $\vec{\theta}_{ij}^{\rm obs}$ and $\vec{\theta}_{ij}^{\rm pred}$ are the observed and the predicted positions for image $i$ in image system $j$, respectively. The predicted image positions $\vec{\theta}_{ij}^{\rm pred}$ are computed given the lens mass parameters $\boldsymbol{\eta}$ and the modelled source position $\vec{\beta}_{j}^{\rm mod}$ of system $j$.  We obtain $\vec{\beta}_{j}^{\rm mod}$ by ray tracing the multiple lensed image positions in system $j$ back to the source plane through the lens equation, and taking the magnification-weighted average of these corresponding source positions.  The rotation matrix $\mathsf{R}$, its transpose $\mathsf{R}^{\rm T}$ and covariance matrix $\mathsf{C_{im}}$ of the image positions depict elliptical positional uncertainties.  Specifically, for an error ellipse of an image position $i$ of the $j^{\rm th}$ multiple image system with semi-major and semi-minor axes of $\sigma_{ij}^{\rm major}$ and $\sigma_{ij}^{\rm minor}$, where the axis $\sigma_{ij}^{\rm major}$ is rotated counter-clockwise from the positive $y$-axis by the position angle $\phi^{\rm im}_{ij}$, the matrices are expressed by
\begin{equation}
\mathsf{R} = \left(\begin{array}{cc} \cos{\phi^{\rm im}_{ij}} & -\sin{\phi^{\rm im}_{ij}} \\ \sin{\phi^{\rm im}_{ij}} & \phantom{-}\cos{\phi^{\rm im}_{ij}} \end{array}\right),
\end{equation}
and
\begin{equation}
\mathsf{C_{\rm im}} = \left(\begin{array}{cc} \left(\sigma^{\rm minor}_{ij}\right)^2 & 0 \\ 0 & \left(\sigma^{\rm major}_{ij}\right)^2 \end{array}\right).
\end{equation}

For the special case where we have circular positional uncertainties with $\sigma^{\rm major}_{ij} = \sigma^{\rm minor}_{ij} = \sigma_{ij}^\text{circ}$, the matrix product ${\mathsf{R}}{\mathsf{C_{im}}}^{-1}{\mathsf{R}}^{\rm T}$ is the identity matrix multiplied by $\sigma_{ij}^{-2}$, and Eq.~(\ref{eq:chi2im_ell}) reduces to the familiar
\begin{equation}
        \chi^2_{\rm im, circ} = \sum_{j=1}^{N_{\rm sys}}\sum_{i=1}^{N_{\rm im}^{j}}\frac{|\Delta \vec{\theta}_{ij}|^2}{\sigma_{ij}^2}.
\end{equation}

To sample the parameter space, we initially used \texttt{EMCEE}, an ensemble sampler using parallel runs of Markov chain Monte Carlo (MCMC) chains \citep{ForemanMackey.2013} to obtain a sampling covariance matrix for more efficient sampling and optimization with simulated annealing \citep{Kirkpatrick1983} and Metropolis-Hastings MCMC chains.

For each mass model in Table~\ref{tab:mass_models}, we ran multiple iterations of Metropolis-Hastings MCMC sampling.  At each iteration, we updated the sampling covariance matrix based on the posterior distribution of the previous iteration.  We adjusted the step size of our MCMC sampling to obtain an acceptance rate of $\sim$0.25, following \citet{Dunkley+05} for efficient sampling.  In our final iteration, we ran two MCMC chains of $2\times10^6$ steps each, where the two chains had different random number seeds.  The two separate chains enabled us to test the statistical accuracy of our results.  

In total, we have 23 multiple image positions from eight systems in Table~\ref{tab:imagepos}, giving us 46 constraints from both the $x$ and $y$ directions.  Furthermore, we have three constraints on the Einstein radii of three galaxies (BCG, cluster member with ID$_{\rm phot}$ of 116, and background LOS galaxy) in our mass model based on their measured velocity dispersions; the Einstein radii of these three galaxies cannot be reconstructed well from the multiple image positions and are mostly constrained by their measured velocity dispersions.  We have a prior on the reference velocity dispersion in the Faber-Jackson relation, but we do not count this as an additional constraint since the prior range is large and this reference velocity dispersion is determined mostly by the multiple image positions.   The number of mass parameters, $N_{\rm par}$, for each of our mass models ranges from 25 to 27, as tabulated in Table~\ref{tab:mass_models}.  The eight multiple image systems additionally require eight source positions, thus 16 $x$ and $y$ values.  Therefore, the number of degrees of freedom (DOF) is DOF = $46 + 3 - N_{\rm par} - 16 = 33 - N_{\rm par}$, which ranges from 6 to 8, depending on the mass model.

The SN host galaxy is lensed into giant arcs covering thousands of image intensity pixels, which provide even more constraints on the cluster mass model.  However, modelling such giant arcs is computationally challenging \citep[e.g.][]{Acebron+24}, and we defer this to future work.

\section{Modelling results and predictions for fixed background cosmology}
\label{sec:results_fixed_cosmo}

In this section we present the mass modelling results for the flat $\Lambda$CDM cosmological model with $H_0=70\, \rm{\, km\, s^{-1}\, Mpc^{-1}}$ and $\Omega_{\rm m} = 1 - \Omega_{\Lambda}=0.3$.  By fixing the background cosmological model, we can quantify uncertainties associated with the mass modelling before using our mass model for cosmography.

We first present the goodness of fit of the seven \GLEE\ mass models and their generic properties in terms of the cluster mass distribution in Sect.~\ref{sec:results_fixed_cosmo:goodness_of_fit}.  We then present our ultimate model in Sect.~\ref{sec:results_fixed_cosmo:best} as a combination of  multiple mass model parameterizations that fit the data comparably well. In Sect.~\ref{sec:results_fixed_cosmo:baseline}, we present our \texttt{baseline} model for direct model comparison with other modelling teams. In Sect.~\ref{sec:results_fixed_cosmo:SNe_predictions}, we predict the image positions and magnifications for both SN Encore and SN Requiem from our ultimate model.  We refrain from presenting the predicted time delays of SN Encore from our model in this paper in order to keep our cosmological analysis blinded. By keeping our time-delay predictions hidden at this stage from the collaborators (Pierel et al. in prep.) who are currently measuring the time delays, their time-delay measurements will not be influenced by our model predictions. The time-delay predictions will be presented in Suyu et al. (in prep.), which will remain hidden from the time-delay analysis group until the time-delay measurements are finalized.  In Sect.~\ref{sec:results_fixed_cosmo:kappa_gamma}, we present the convergence and shear values at the SN Encore image positions for future microlensing studies.

\subsection{Goodness of fit of the mass models}
\label{sec:results_fixed_cosmo:goodness_of_fit}

\begin{figure*}[!htb]
\centering

    \begin{subfigure}[!htb]{0.32\textwidth}
        \centering
        \includegraphics[width=\textwidth]{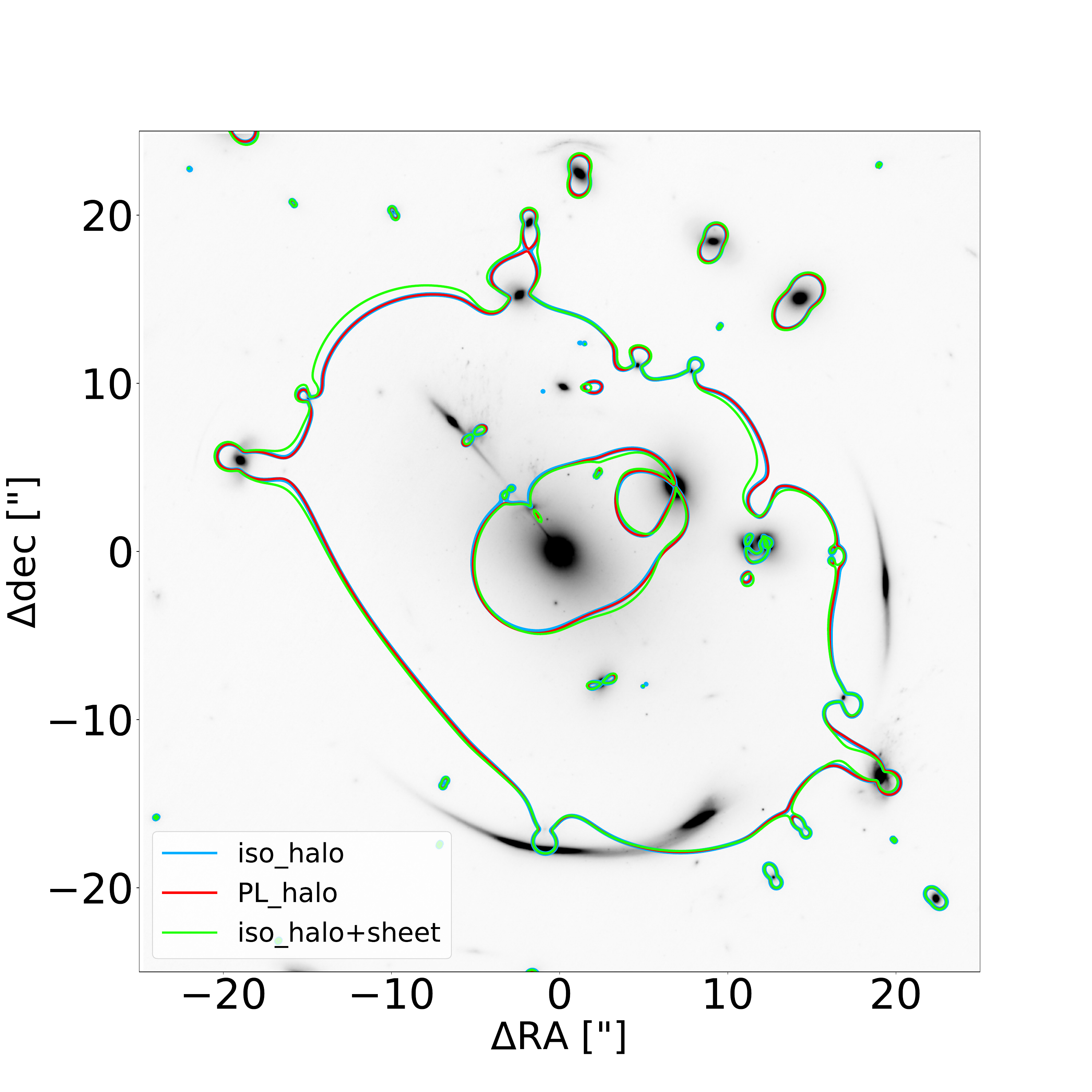}
    \end{subfigure}
\hfill
    \begin{subfigure}[!htb]{0.32\textwidth}
        \centering
        \includegraphics[width=\textwidth]{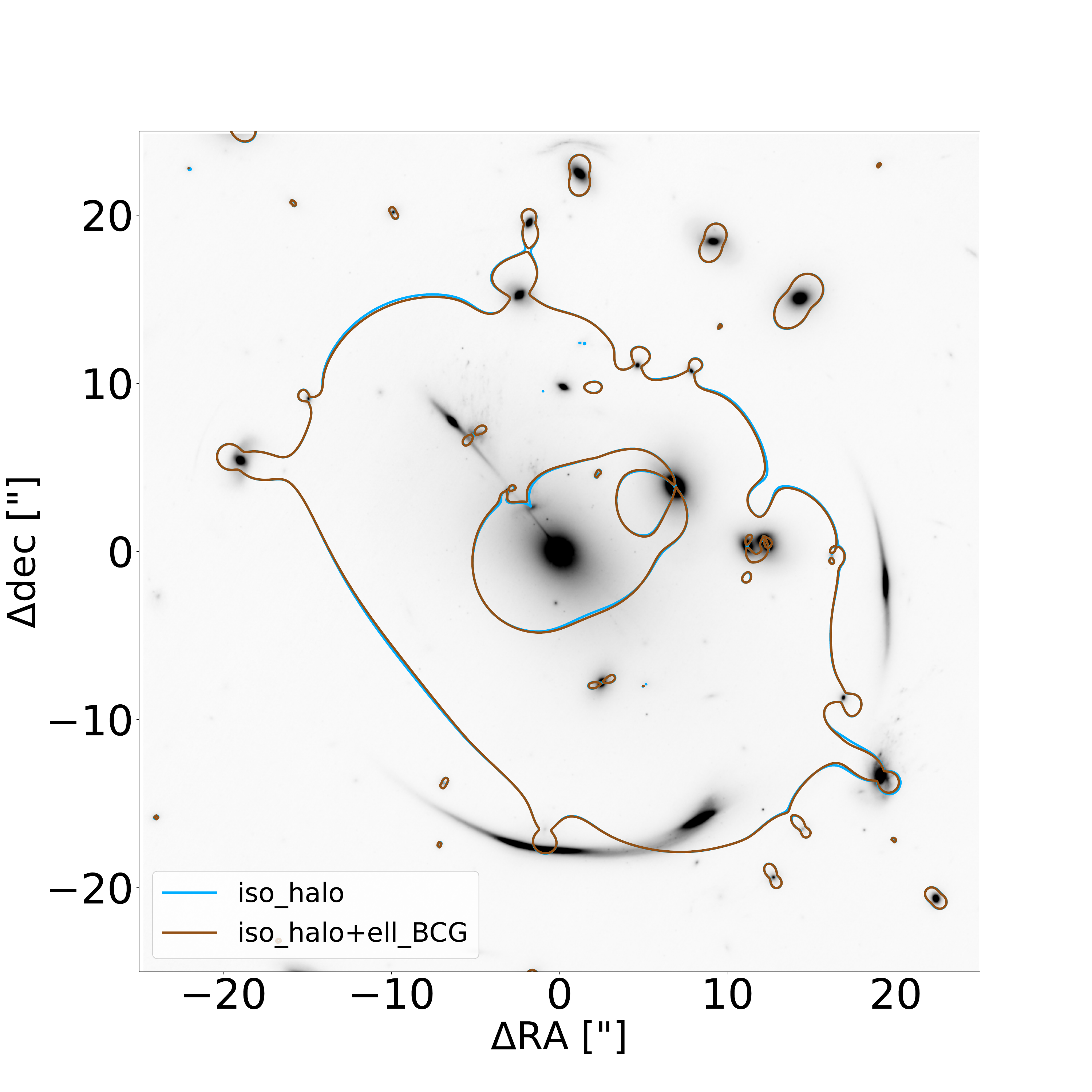}
    \end{subfigure}
\hfill
    \begin{subfigure}[!htb]{0.32\textwidth}
        \centering
        \includegraphics[width=\textwidth]{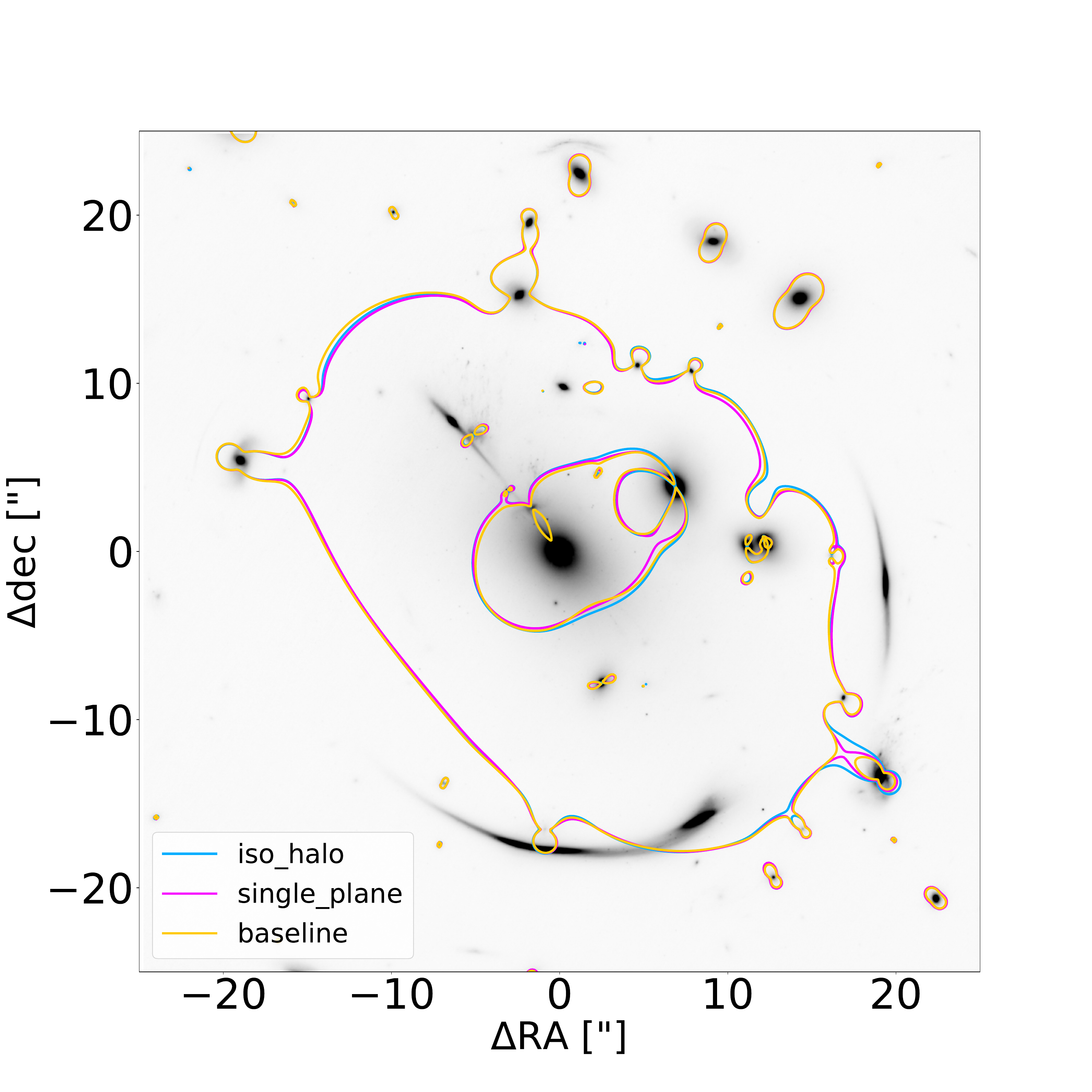}
    \end{subfigure}
    
        \caption{Critical curves of the multiple models for a source at redshift $z=1.949$, which is the redshift of SN Encore (and SN Requiem). The panels show the \texttt{iso\_halo} model in blue (all panels), \texttt{PL\_halo} in red (left panel),  \texttt{iso\_halo+sheet} in green (left panel), \texttt{iso\_halo+ell\_BCG} in brown (middle pannel), \texttt{single\_plane} in magenta (right panel), and \texttt{baseline} in orange (right panel). The best-fit model was used for each curve. The critical curves of the different models are overall similar, especially near the southern giant arc and the radial arc (of the lensed host galaxy of SN Encore), and hence the overlapping critical lines.}
        \label{fig:crit_curves}
\end{figure*}

In Table \ref{tab:mass_models}, we list the 
$\chi^2$ value of the most probable (highest posterior) mass distribution of each model in fitting to the multiple image positions.  All models apart from the \texttt{NFW\_halo} have the most-probable $\chi_{\rm im,MP}^2 \lesssim 7$.  Given that the corresponding DOF are between 6 and 8, we achieved a reduced $\chi^2 \sim 1$ for these models, showing a good fit to the observables.  The \texttt{NFW\_halo} model has its most-probable $\chi_{\rm im,MP}^2$ of 23 that is substantially higher than all other models, demonstrating that the cluster main dark matter halo is better represented by a cored isothermal profile.  Given this result, we did not further consider the \texttt{NFW\_halo} model in this work.

When comparing the mass model parameter constraints and especially the predicted SN image positions, magnifications and time delays between the two MCMC chains, we find good agreement, with the model predicted values typically agreeing within $0.5\sigma$.  Given the large number of mass model parameters, including a few parameters such as the truncation radii that are poorly constrained, the sampling of the full posterior distribution needs a long MCMC chain.  This is the reason why the two chains of $2\times10^6$ in length do not provide perfectly matching parameter constraints.  

Focusing specifically on the predicted time delay between images SN1a and SN1b of SN Encore (see Fig.~\ref{fig:imgpos}) for cosmographic analysis based on this time-delay pair (with anticipated time-delay measurement in the near future), the offset of the median time delay values between the two chains is within $\sim1\%$ for half of the six models (not considering the \texttt{NFW\_halo} model), although the offset could reach up to $\sim4\%$ in the cases of the \texttt{PL\_halo} and \texttt{baseline} models.  Since we anticipate that the upcoming time-delay measurement uncertainty will be $\gtrsim 5\%$ based on existing data, our parameter uncertainties of $\lesssim 4\%$ is currently tolerable.  In contrast to the shortest time delay between SN1a and SN1b, the longer time delays of the other lensed images have more accurate predictions; all models have $<1\%$ difference between the two chains for the delay pair of SN1a and (future) SN1d.  Similarly, the model predictions are accurate to within 1\% for the corresponding pair (2d-2a) of SN Requiem images.   Therefore, with a future time-delay measurement of the reappearance of SN Requiem or SN Encore, the uncertainty due to statistical fluctuations in the sampling of our model is subpercent.  

Given the overall consistency in the two chains for each mass model, we picked one of the two chains to present our results.  All six models (i.e., excluding the \texttt{NFW\_halo} model that does not fit well to the observed data) show similar global properties for the galaxy cluster, with a concentrated primary DM halo centred within $3\arcsec$ of the BCG and an extended perturbative DM halo that is located $\sim30\arcsec$ southwest of the BCG in a region with cluster members (Fig.~\ref{fig:model_galaxies}).  The perturbative DM halo has a large core radius of $\sim30\arcsec$ for the four multi-lens-plane models, whereas it has a core radius of $\sim20\arcsec$ for the \texttt{single\_plane} model and $\sim50\arcsec$ for the \texttt{baseline} model.  The cluster halo is dominated by the primary halo component, and the perturbative halo adds more mass near the region of the giant tangential arcs, making the overall cluster halo non-elliptical.  This finding is in agreement with the independent mass model of \citet{Acebron+25}, who showed that the perturbation likely originates from the mass concentration located $\sim140\arcsec$ southeast of the BCG.  The need for such additional perturbative halo components to go beyond elliptical haloes (which may be overly simplistic) and fit strong-lensing constraints has also been found in studies of other clusters \citep[e.g.][]{Limousin+25}.

The modelled source positions of image systems 4.1 and 4.3 (that are identified from the MUSE data, see Table \ref{tab:imagepos} and Fig.~\ref{fig:imgpos}) of the most probable models are closely located on the source plane of the OII emitter.  For the \texttt{iso\_halo+sheet} model, the source positions of systems 4.1 and 4.3 are located within 0.11\arcsec; for the other five models (\texttt{iso\_halo}, \texttt{PL\_halo}, \texttt{iso\_halo+ell\_BCG}, \texttt{single\_plane}, \texttt{baseline}), the source positions are within 0.05\arcsec.  Therefore, even though systems 4.1 and 4.3 were treated as two separate image systems of four and two multiple images, respectively, they are in fact likely a single source system with six images.  The treatment of a 6-image system as two separate 4-image and 2-image systems would provide two fewer constraints on the mass model, but would not lead to biases on the mass model parameter values and $H_0$.  A more detailed study of the systems 4.1 and 4.3 is deferred to future work.

\subsection{Ultimate model}
\label{sec:results_fixed_cosmo:best}

The four multi-plane lens models, \texttt{iso\_halo}, \texttt{PL\_halo}, \texttt{iso\_halo+sheet,} and \texttt{iso\_halo+ell\_BCG}, all fit well the observed image positions with an image $\chi_{\rm im,MP}^2 \leq 6.3$, and a reduced $\chi_{\rm im,MP}^2 \sim 1$.  The different parameterizations enable us to incorporate systematic uncertainties from model assumptions.  Despite the different mass parameterizations and even considering a model that incorporates explicitly a mass sheet to account for the  mass-sheet degeneracy (which is expected to be the dominant mass-modelling uncertainty), we obtained globally similar cluster mass properties.  The critical curves for the different models shown in Fig.~\ref{fig:crit_curves} (left and middle panels) are nearly identical for these four mass models.  The predicted magnifications and time delays of both SNe Requiem and Encore from the four different models mostly agree within their estimated 1$\sigma$ uncertainties, although for a few cases the magnification and delay predictions from the different models (obtained from samples of the final MCMC chain, which we describe in more detail in Sect.~\ref{sec:results_fixed_cosmo:SNe_predictions}) are more discrepant but still within 2$\sigma$.

We list the parameters of each model and their inferred values in Table \ref{tab:encore_final_param}.  The parameter values for the galaxies (cluster members, foreground and background galaxies) agree within their $1\sigma$ uncertainties across the four models (\texttt{iso\_halo}, \texttt{PL\_halo}, \texttt{iso\_halo+sheet} and \texttt{iso\_halo+ell\_BCG} in columns 4 to 7).  The reference galaxy has a lower velocity dispersion ($\sim170-180\, {\rm km\,s^{-1}}$) compared to the prior ($206\pm25\, {\rm km\,s^{-1}}$), indicating that the cluster mass models require on average slightly lower velocity dispersions for the cluster members compared to the Faber-Jackson fit, although the preferred values from the lens mass model are consistent within $\sim$$1\sigma$ of the Faber-Jackson relation.  In the case of the \texttt{iso\_halo+ell\_BCG} where the BCG is elliptical (instead of circular as in other models), the axis ratio of the BCG is $\sim$$0.9$ which is nearly circular, and thus its position angle has broad uncertainties.  This shows that a circular BCG mass distribution is a good approximation for fitting to the current data of multiple image positions. The primary DM halo parameters also agree well across the four different models, and its position angle is consistent with the observed position angle of the BCG light profile.  The \texttt{PL\_halo} model with a radially variable DM density profile has a resulting radial profile slope of $\gamma=0.51\pm0.03$, which agrees well with the other isothermal halo models (with $\gamma=0.5$) within the uncertainties.  The centroid of the primary DM halo is also closely aligned with that of the BCG (within $\sim$$1.5\arcsec$).  In contrast, the perturbative DM halo centroid is $\sim$30\arcsec\ southwest of the BCG, and has a large core radius.  Furthermore, there is degeneracy in the perturbative halo component and the mass sheet in the \texttt{iso\_halo+sheet} model, where the slightly positive convergence sheet of $\sim$$0.09$ contributes to the extended DM halo, resulting in a smaller Einstein radius of the perturbative DM halo compared to the other three mass models.  In summary, Table \ref{tab:encore_final_param} shows a high level of consistency across the different models, despite the different mass-model parameterizations.

As an illustration, we show a breakdown of the contributions to $\chi_{\rm im,MP}^2$ from each multiple image system for the \texttt{iso\_halo} model.  For the SN Encore (system 1), SN Requiem (system 2), and their host galaxy centroid (system 3), we are able to reproduce their positions within a root-mean-square (rms) scatter of $\lesssim$$0.03\arcsec$ for each system, which is comparable to the positional uncertainties.  We are also able to reproduce the image positions of system 4.2 ([OII] arc with positions from the JWST image) with a rms of $0.02\arcsec$. The remaining image systems have larger rms offsets between the predicted and observed image positions, ranging from $0.21\arcsec$ to $0.44\arcsec$, because of the larger observed positional uncertainties from the MUSE data (as listed in Table \ref{tab:imagepos}).  Overall, we fit all 23 identified multiple image positions within their observational uncertainties and with an rms scatter of $0.24\arcsec$ for the \texttt{iso\_halo} model.  The results for the \texttt{PL\_halo}, \texttt{iso\_halo+sheet} and \texttt{iso\_halo+ell\_BCG} models are similar, given their similar values of $\chi^2_{\rm im,MP}$.

\begin{figure}
        \centering
        \includegraphics[width=\linewidth]{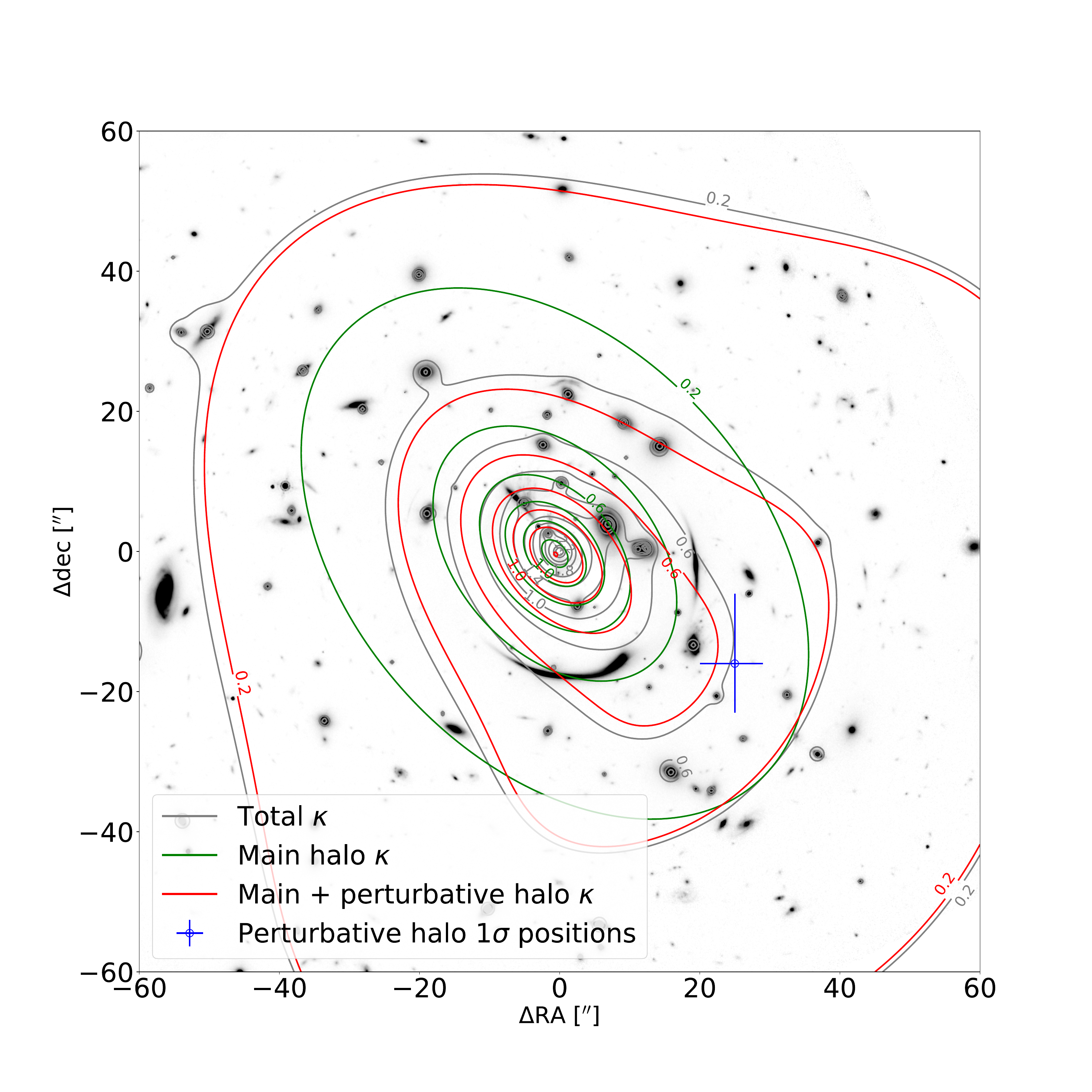}
        \caption{Effective convergence, $\kappa_{\rm eff}$, of the \texttt{iso\_halo} model for the SN Encore redshift ($z=1.949$).  The total $\kappa_{\rm eff}$ distribution (in grey) near the cluster core (inner $\sim$10\arcsec\ region) is dominated by the primary DM halo and the BCG, and $\kappa_{\rm eff}$ follows roughly the observed BCG and galaxy intensity distribution.  The extension of the contour level with $\kappa_{\rm eff}=0.6$ towards the southwest part of the BCG is due to the presence of the perturbative DM halo located around $(x_{\rm DM2},y_{\rm DM2})\sim(20\arcsec,-20\arcsec)$, for which we show its position in blue with its 1$\sigma$ uncertainty. The green contours show the $\kappa_{\rm eff}$ of only the main DM halo, the red contours show $\kappa_{\rm eff}$ for the main and perturbative halo combined, which highlights the deformation of the overall $\kappa_{\rm eff}$ due to the perturbative halo.}
        \label{fig:kappa}
\end{figure}

In Fig.~\ref{fig:kappa}, we show the effective convergence (dimensionless surface mass density) for the SN redshift ($z=1.949$) from our multi-lens-plane \texttt{iso\_halo} model that accounts for the three lens redshift planes ($z_{\rm fg}=0.309$ of the foreground galaxy, $z_{\rm d}=0.336$ of the galaxy cluster, and $z_{\rm bg}=0.371$ of the background galaxy).  The effective convergence, $\kappa_{\rm eff}$, is obtained by taking $\frac{1}{2}\boldsymbol{\nabla} \cdot \boldsymbol{\alpha}_{\rm tot}$, where $\boldsymbol{\alpha}_{\rm tot}$ is the total (scaled) deflection angle from summing up the deflection angles $\boldsymbol{\hat{\alpha}}$ in all lens planes to the $z=1.949$ source plane: $\boldsymbol{\alpha}_{\rm tot} = \sum_{i=1}^{N-1} \frac{D_{iN}}{D_N}\boldsymbol{\hat{\alpha}(\theta}_i)$ with the source plane as the $N^{\rm th}$ plane \citep[e.g.][]{Gavazzi08, Chirivi+18}.  For our specific case with three lens redshift planes, the source plane is thus at $N=4$. 
The contours of the effective convergence show that most of the mass is in the primary dark matter halo given the nearly concentric elliptical contours.  The contours broaden slightly beyond the tangential arcs in the southwest direction from the BCG, due to the presence of the perturbative dark matter halo.  Despite the seemingly small amount of convergence associated with the perturbative halo, its presence is necessary to fit to the observed image positions with a reduced $\chi_{\rm im}^2 \sim 1.$  Without this perturbative halo, the $\chi_{\rm im}^2$ was an order of magnitude higher.

We obtained our ultimate model based on the lensed image positions by combining these four multi-lens-plane models (\texttt{iso\_halo}, \texttt{PL\_halo}, \texttt{iso\_halo+sheet,} and \texttt{iso\_halo+ell\_BCG}) with equal weights.  Our ultimate model therefore incorporates systematic uncertainties due to mass model parameterizations in predicting the properties (e.g. positions, magnifications and time delays) of SN Encore and SN Requiem, and also for measuring cosmological parameters in future studies.

\subsection{\texttt{single\_plane} and \texttt{baseline} models}
\label{sec:results_fixed_cosmo:baseline}

We consider a scenario where we place the LOS galaxies at the same redshift as the galaxy cluster.  This will enable us to assess the impact of assuming a single lens plane in the mass modelling of the MACS J0138$-$2155 cluster, especially since most of the cluster-lens models to date adopt a single lens plane.  Furthermore, the positional uncertainties of multiple lensed images are often considered to be circular rather than elliptical.  Circular uncertainties are sufficient approximations especially when the rms offset between the predicted and observed image positions is substantially larger than the nominal positional uncertainties, signaling that additional origins of uncertainties associated with mass model assumptions dominate over the positional uncertainty.  Since several of our mass models in Table \ref{tab:mass_models} fit well to the observed image positions within the elliptical positional uncertainties, such that the usages of elliptical versus circular uncertainties could lead to differences, we assess here the difference in modelling results between elliptical and circular positional uncertainties.  

We thus consider two specific models: (1) the \texttt{single\_plane} model, which takes the \texttt{iso\_halo} model and assumes the foreground galaxy at $z_{\rm fg}=0.309$ and background galaxy at $z_{\rm bg}=0.371$ are both at the cluster redshift $z_{\rm d}=0.336$ instead; and (2) the \texttt{baseline} model, which is the same as the \texttt{single\_plane} model but with circular positional uncertainties $\sigma^{\rm circ}$ that is obtained as the geometric mean of the semi-major and semi-minor axes of the elliptical positional uncertainty, i.e., $\sigma^{\rm circ} = \sqrt{\sigma^{\rm major} \sigma^{\rm minor}}$.  As noted in Sect.~\ref{sec:mass_model:models}, the \texttt{baseline} model with a single lens plane and circular positional uncertainties is the setup that all the seven independent modelling teams in Suyu et al. (in prep.) have agreed to construct in order to facilitate a direct model comparison.

We show the critical curves of the \texttt{single\_plane} (magenta) and \texttt{baseline} (orange) models in comparison to that of the \texttt{iso\_halo} (blue) model in the right-hand panel of Fig.~\ref{fig:crit_curves}.  The critical curves again overlap well, even at the locations near the foreground galaxy (at $(x_{\rm fg},y_{\rm fg})\sim (-1\arcsec,-17\arcsec)$) and the background galaxy (at $(x_{\rm bkg},y_{\rm bkg})\sim(7\arcsec,4\arcsec)$) that are approximated at the cluster redshift in the \texttt{single\_plane} and \texttt{baseline} models.  There is a noticeable difference in the critical curves near the jellyfish galaxy at $(x_{\rm JF1},y_{\rm JF1})\sim(19\arcsec,-13\arcsec)$ where the \texttt{single\_plane} and \texttt{baseline} models have a lower resulting Einstein radius for this jellyfish galaxy.  Nonetheless, since the multiple lensed image positions and the giant arcs are not near this critical curve, the difference in the critical curve is expected to have minimal impact on the model properties of SN Encore and SN Requiem.

We tabulate the model parameter values of the \texttt{baseline} model in Table \ref{tab:encore_final_param} in the last column.  In comparison to the values of \texttt{iso\_halo}, we see that almost all the parameter values agree within their $1\sigma$ uncertainties, except for the Einstein radius of the first jellyfish galaxy that is almost $\sim2\sigma$ lower in the \texttt{baseline} model.  We find that adopting single-plane lensing with elliptical positional uncertainties, i.e. the \texttt{single\_plane} model, shifts the predicted magnification and time delays by $\sim1\sigma$ for most predictions, and up to $\sim2\sigma$ for a few predictions, relative to the \texttt{iso\_halo} model.  Curiously, the change of elliptical to circular positional uncertainties in the \texttt{baseline} reduces the shift relative to the \texttt{iso\_halo} to within $\sim 1\sigma$.  While it is merely a coincidence that the biases induced by the two assumptions of single-plane lens and circular-positional uncertainties cancel each other to some extent, this finding suggests that the \texttt{baseline} model provides an approximation of the galaxy cluster MACS J0138$-$2155 mass distribution that is accurate to $\sim 1\sigma$ of the mass modelling uncertainties in comparison to our ultimate model in Sect. \ref{sec:results_fixed_cosmo:best}.  In Table \ref{tab:imgchi2}, we show in columns 4 and 5 the breakdown of $\chi^2_{\rm im,MP}$ and rms scatter for the most-probable \texttt{baseline} model.  The values are  similar to those of the \texttt{iso\_halo} model, with the same resulting rms scatter of 0.24\arcsec\ between all the observed and predicted image positions.

\begin{table}[H]
  \caption[Image chi-square of systems.]{Root-mean-square (rms) offset and $\chi^2_{\rm im,MP}$ between the observed and model-predicted image positions from the \texttt{iso\_halo} and \texttt{baseline} models.}
  \label{tab:imgchi2}
  \begin{center}
  \begin{tabular}{llllll}
    \toprule
     Image  & \multicolumn{2}{c}{\texttt{iso\_halo}} & & \multicolumn{2}{c}{\texttt{baseline}} \\
     \addlinespace[2pt]
        \cline{2-3} \cline{5-6} \addlinespace[2pt]
     system  & $\chi^2_{\rm im}$ & rms & & $\chi^2_{\rm im}$ & rms \\
    \midrule

1 &  0.003 & 0.0014\arcsec & & 0.08 & 0.008\arcsec   \\ 
2 &  0.46 & 0.031\arcsec & & 0.16 &  0.018\arcsec \\ 
3 &  0.34 & 0.014\arcsec & & 0.88 & 0.025\arcsec \\ 
4.1 & 2.21 & 0.44\arcsec & & 3.02 & 0.46\arcsec \\ 
4.2 & 0.07 & 0.020\arcsec & & 0.13 & 0.032\arcsec \\ 
4.3 & 0.30 & 0.21\arcsec & & 0.38 & 0.19\arcsec \\ 
5 &  1.87 & 0.33\arcsec & & 1.45 & 0.27\arcsec \\ 
6 &  1.03 & 0.29\arcsec & & 1.24 & 0.32\arcsec \\ 
\midrule  
all & 6.27 & 0.24\arcsec & & 7.34 & 0.24\arcsec \\
\bottomrule

\end{tabular}
  \end{center}
  \end{table}

\subsection{Predicted positions and magnifications of SN Encore and SN Requiem}
\label{sec:results_fixed_cosmo:SNe_predictions}

To predict the multiple image positions of a given model, we first map the individual observed image positions back to the source plane through the multi-plane lens equation. We then obtain the model source position as the magnification-weighted average of the mapped source positions.  Given the model source position, we then solve the non-linear lens equation to obtain the model-predicted image positions.  We also compute the lensing magnifications at the model-predicted image positions.

For our ultimate model, which is the combination of the four multi-plane mass models, we combined the respective chains, each of length $2\times10^6$, to obtain a final chain length of $8\times10^6$.  We then thinned this chain by a factor of 1000, reaching a final thinned chain of 8000 samples.  For each sample in this chain, we then solved the lens equation to obtain the predicted image positions and magnifications.  For the \texttt{baseline} model, we similarly thinned it by a factor of 1000 and use the thinned chain to predict the image positions/magnifications.  

We find that all samples in the chains predict at least four images for either SN Encore and SN Requiem.  For SN Encore, the four predicted images are near the 1a, 1b, 1c and 1d locations marked on Fig.~\ref{fig:color_img}.  For SN Requiem, the four images that are persistently predicted by the models are similar in configuration to that of SN Encore; we denote them as images 2a, 2b, 2c (matching their corresponding observed images) and 2d.  Images 1d and 2d are predicted near the radial arc, and will appear in the future given their model time delays of multiple years relative to the first images 1a and 2a, respectively.  While some mass models in the MCMC chain predict a fifth central image of SN Encore or SN Requiem near the core of the BCG, this image is not always predicted.  We denote these images as 1e for SN Encore and 2e for SN Requiem, and compute the percentage of models in the chains that predict these fifth images.  Furthermore, some samples in the chain produce additional (future) images of SNe Encore and Requiem near the jellyfish galaxies along the radial arc.  Similarly, some samples in the chain produce additional images of SN Encore formed by the foreground galaxy (at ($x_{\rm fg}, y_{\rm fg}$) listed in Table \ref{tab:encore_final_param}) near image 1b.  Since these additional images are not persistently predicted across all samples in the chain and are sensitive to the mass distributions of the jellyfish galaxies or the foreground galaxy, which are simplistically approximated as spherical in our model, we defer to future work for investigating in detail these non-persistent images.  Future detections of such additional images from the jellyfish galaxies, if any, will help place additional constraints on the mass distribution of inner cluster core and the jellyfish galaxies.

In Table \ref{tab:SNpred:best} we list the predicted image positions and magnifications of both SN Encore and SN Requiem based on our ultimate model.  The predicted positions of 1a, 1b, 2a, 2b, and 2c agree well with the observed positions in Table \ref{tab:imagepos} within both the observational uncertaintes in Table \ref{tab:imagepos} and the model uncertainties in Table \ref{tab:SNpred:best}, which are in line with the image position rms and $\chi_{\rm im}^2$ computed in Table \ref{tab:imgchi2}. The model positional uncertainties on these image positions are on the order of tens of milliarcseconds, as shown in Table \ref{tab:SNpred:best}.  The model positional uncertainties of undetected/future images (1c, 1d, 1e, 2d, 2e) are larger, up to $\sim$0.7$\arcsec$.

We list the predictions from the \texttt{baseline} model in Table \ref{tab:SNpred:baseline}. The predictions from the ultimate model and the \texttt{baseline} models agree well overall within $\sim$1$\sigma$ of the model uncertainties.  This shows that the \texttt{baseline} model is a sufficient approximation for modelling SN Encore and SN Requiem, based on the current image positions and kinematic constraints.  Nonetheless, we caution that there is a larger shift in the model predictions when comparing the \texttt{single\_plane} model (with elliptical positional uncertainties) to the ultimate model.  Therefore, to err on the cautious side, we advocate for using the ultimate model from multi-plane lensing for future cosmographic analysis, particularly in measuring the value of $H_0$.

The magnification values of the multiple images predicted from both our ultimate and \texttt{baseline} models are substantially higher than the values predicted by \citet{Newman2018a} and \citet{Rodney2021} for the SN host galaxy and SN Requiem, by a factor up to $\sim$6.  In contrast, our predicted magnifications of SN Requiem agree with those of \citet{Acebron+25} within the estimated uncertainties, and we refer to \citet{Acebron+25} for a more detailed discussion.
\

\begin{table}[h!]
    \centering
    \caption{Predicted image positions and magnifications of SN Encore and SN Requiem from the ultimate lens mass model.}
    \begin{tabular}{lccc}
        \toprule
        image & $x$ coordinate $[\arcsec]$ & $y$ coordinate $[\arcsec]$ & magnification \\ 
        \midrule
        \midrule
        \multicolumn{4}{l}{SN Encore predictions} \\
        \midrule
        1a & $\phantom{-}8.40\pm0.04$ & $-16.07\pm0.01$  & $-28.0^{+2.5}_{-3.7}$ \vspace{2px} \\ 
        1b & $\phantom{-}0.23\pm{0.04}$ & $-17.86\pm{0.01}$     & $\phantom{-}40.0^{+7.6}_{-6.3}$ \vspace{2px} \\
        1c* & $\phantom{-}19.43\pm{0.03}$ & $-2.62\pm{0.13}$    & $\phantom{-}12.1^{+1.0}_{-0.9}$ \vspace{2px} \\
        1d* & $-6.18^{+0.10}_{-0.09}$ & $\phantom{-}7.63\pm{0.09}$      & $-2.3^{+0.3}_{-0.4}$ \vspace{2px} \\ 
        1e*$^{\dagger}$ & $-0.88^{+0.19}_{-0.50}$ & $\phantom{-}1.23^{+0.70}_{-0.27}$   & \phantom{-}$2.8^{+4.3}_{-1.3}$ \vspace{2px} \\ 
        \midrule
        \midrule
        \multicolumn{4}{l}{SN Requiem predictions} \\
        \midrule
        2a & $11.30\pm{0.07}$ & $-15.50\pm{0.04}$ & $-39.8^{+5.4}_{-6.5}$ \vspace{2px} \\
        2b & $1.99^{+0.08}_{-0.07}$ & $-18.61\pm{0.02}$ & $\phantom{-}24.3^{+2.9}_{-2.2}$ \vspace{2px} \\
        2c & $18.84\pm{0.03}$ & $-6.65\pm{0.07}$ & $\phantom{-}15.7^{+1.9}_{-1.5}$ \vspace{2px} \\
        2d* & $-4.17^{+0.22}_{-0.21}$ & $5.64^{+0.29}_{-0.30}$ & $-2.2^{+0.4}_{-0.5}$ \vspace{2px} \\
        2e*$^\dagger$ & $-1.63^{+0.20}_{-0.03}$ & $2.41^{+0.14}_{-0.32}$ & \phantom{-}$1.6^{+0.8}_{-1.2}$ \vspace{2px} \\
        \bottomrule
        \vspace{1px} 
    \end{tabular}
    \label{tab:SNpred:best}
    \tablefoot{The coordinates are relative to the BCG that is centred at $(x_{\rm BCG},y_{\rm BCG})=(0,0)$.  Lensed images marked by an asterisk (*) are model predictions that have not yet been clearly detected, either due to their faintness (image 1c) or due to their long time delays (images 1d, 1e, 2d, and 2e), and are therefore not included in the model as constraints (see Table~\ref{tab:imagepos}). Lensed images marked by a cross ($^{\dagger}$ ) are not predicted by all the models in the MCMC chain.  Image 1e is predicted by 35\% of the models in the chain and Image 2e by only 3\% of the models in the chain; all the other images are predicted by all the models in the chain.  }
\end{table}

\subsection{Convergence and shear of SN Encore}
\label{sec:results_fixed_cosmo:kappa_gamma}

To enable future microlensing studies of SN Encore, we list in Table \ref{tab:kappa_gamma} the effective convergence and shear values at the positions of SN Encore multiple images from the four multi-plane models that constituted the ultimate model.  For the images 1a and 1b, we evaluated the convergence/shear at the observed image positions (in Table \ref{tab:imagepos}); for image 1c, that is not clearly visible, and the future image 1d, we used the median of the predicted image positions of the ultimate model (in Table~\ref{tab:SNpred:best}).  Since images 1e is not always predicted to be present in our mass models, we do not list its convergence and shear here.

\begin{table}[h!]
    \centering
    \caption{Effective convergence and shear values at the SN Encore multiple image positions.}
    \begin{tabular}{lccc}
        \toprule
        lens model & SN image & {$\kappa$} & {$\gamma$} \\ 
        \midrule
        \texttt{iso\_halo} & 1a & 0.75  & 0.31  \\
                        & 1b & 0.66 & 0.30 \\
                        & 1c & 0.63 & 0.25 \\
                        & 1d & 1.08 & 0.62 \\
        \midrule
        \texttt{PL\_halo} & 1a & 0.74 & 0.31 \\
                        & 1b & 0.65 & 0.31 \\
                        & 1c & 0.62 & 0.25 \\
                        & 1d & 1.08 & 0.63 \\
        \midrule
        \texttt{iso\_halo+sheet} & 1a & 0.73 & 0.33 \\
                        & 1b & 0.64 & 0.31 \\
                        & 1c & 0.61 & 0.26 \\
                        & 1d & 1.09 & 0.59 \\
        \midrule
        \texttt{iso\_halo+ell\_BCG} & 1a & 0.75 & 0.30 \\
                        & 1b & 0.66 & 0.30 \\
                        & 1c & 0.63 & 0.25 \\
                        & 1d & 1.08 & 0.62 \\
        \bottomrule
    \end{tabular}
    \label{tab:kappa_gamma}
\end{table}

The most probable convergence and shear values from the four models mostly agree within 0.02 and all within 0.04, showing broad consistency in the results from the different mass model parameterizations.  For microlensing studies, the differences in convergence and shear across the models are small in comparison to the less well-constrained smooth matter mass fraction (which is the convergence due to the smooth DM distribution in fraction of the total convergence composed of DM and baryons) and also the stochasticity of stellar distributions.  The use of any of the convergence and shear values listed in Table \ref{tab:kappa_gamma} for microlensing studies should therefore suffice.  Microlensing can be significant especially for image 1b, that has the foreground galaxy located within $2''$, and image 1d, that has a jellyfish galaxy stripped across the neighbouring area.  Future microlensing studies of this system should account for the effects of these nearby galaxies.

\section{Summary}
\label{sec:summary}

We constructed new lensing models of the galaxy cluster MACS J0138$-$2155 based on new JWST imaging and MUSE spectroscopy.  The analysis of the spectroscopic dataset is performed by \citet{Granata+24}, which we combined with the photometric analysis here to obtain the key ingredients for building our cluster total mass models.  We summarize the main results as follows.

\begin{itemize}
    \item We fitted nearly 500 galaxies in the FOV of the HST and JWST imaging data using S\'{e}rsic profiles and we obtained their photometries across six JWST and five HST filters. 
    
    \item We used these galaxy photometries to select 34 additional cluster members based on colour-magnitude and colour-colour criteria, which complemented the spectroscopic sample of 50 cluster members and resulted in a total of 84 cluster members.  The photometric catalogue of the cluster members is publicly released along with this paper.
    
    \item In addition to the cluster, there is one foreground galaxy (near the southern giant arc) and one massive background galaxy, which we included explicitly in our multi-lens-plane mass model.
    
    \item We identified the image positions of eight multiple image systems with a total of 23 images from four distinct spectroscopic redshifts in a range of $0.767 \leq z \leq 3.420$, based on the combination of JWST, HST, and MUSE data.
    
    \item We considered seven different lens mass models with \GLEE, exploring a range of DM halo profiles, the inclusion of a mass sheet, the impact of multi-plane versus single-plane modelling, and the effect of having elliptical versus circular BCG.  With two DM halo components (one main DM halo and a second, perturbative component) and most of the cluster galaxies that follow the Faber-Jackson relation, six of the seven models fit to the observed image positions with a reduced $\chi^2_{\rm im} \sim 1$.  Cored-isothermal DM haloes could fit well to the observables, whereas the NFW DM halo results in a reduced $\chi^2_{\rm im}$ that is approximately four times higher.
    
    \item We combined the four multi-lens-plane mass models (\texttt{iso\_halo}, \texttt{PL\_halo}, \texttt{iso\_halo+sheet} and \texttt{iso\_halo+ell\_BCG}, which have remarkably consistent results) to form our ultimate model for predicting observables of SN Encore and SN Requiem, and also for future cosmographic analyses.

    \item We considered a \texttt{baseline} model with the two LOS galaxies approximated at the cluster redshift and with circularized positional uncertainties of the image positions.  The \texttt{baseline} model is designed to facilitate the direct comparison of independent mass models from different teams (Suyu et al. in prep.).  We found the \texttt{baseline} model provides a good approximation to our ultimate model.

    \item We predicted the positions and magnifications of SN Encore and SN Requiem, including ones that will only appear in the future.  Observations of these predicted images in the future will provide a unique opportunity to test our models.

    \item We provide the convergence and shear values at the positions of the SN Encore multiple images to enable future microlensing studies.
    
\end{itemize}

The exciting discovery of SN Encore in the galaxy that previously hosted SN Requiem gives us an excellent opportunity to study cosmology, particularly to constrain the Hubble constant, and the SN host galaxy.  Our work in building an accurate total mass model of the cluster is crucial as it lays the foundation for enabling state-of-the-art cosmological and astrophysical studies of this cluster.  The dataset presented in this paper also forms the basis for independent cluster-mass modelling of this cluster.  The comparison of seven independent mass models based on this unique dataset will be presented in a forthcoming publication.

\begin{dataav}
Table \ref{tab:mag_cat} is available at the CDS via \url{https://cdsarc.cds.unistra.fr/viz-bin/cat/J/A+A/702/A157}.
\end{dataav}

\begin{acknowledgements}
We thank Jake Summers and Jordan D'Silva for their help in reducing the JWST images, and the anonymous referee for the helpful feedback.  
This paper is based in part on observations with the NASA/ESA Hubble Space Telescope and James Webb Space Telescope obtained from the Mikulski Archive for Space Telescopes at STScI, which is operated by the Association of Universities for Research in Astronomy, Inc., under NASA contract NAS 5–26555 for HST, and under NASA contract NAS 5-03127 for JWST. The specific observations used in this work can be accessed via DOI:\href{https://doi.org/10.17909/snj9-an10}{10.17909/snj9-an10}.
SE and SHS thank the Max Planck Society for support through the Max Planck Fellowship for SHS. 
This project has received funding from the European Research Council (ERC) under the European Union's Horizon 2020 research and innovation programme (LENSNOVA: grant agreement No 771776).
This work is supported in part by the Deutsche Forschungsgemeinschaft (DFG, German Research Foundation) under Germany's Excellence Strategy -- EXC-2094 -- 390783311.  
SS has received funding from the European Union’s Horizon 2022 research and innovation programme under the Marie Skłodowska-Curie grant agreement No 101105167 — FASTIDIoUS. We acknowledge financial support through grants PRIN-MIUR 2017WSCC32 and 2020SKSTHZ. 
AA acknowledges financial support through the Beatriz Galindo programme.  AA and JMD acknowledge support from the project PID2022-138896NB-C51 (MCIU/AEI/MINECO/FEDER, UE), Ministerio de Ciencia, Investigaci\'on y Universidades. 
YF is supported by JSPS KAKENHI Grant Numbers JP22K21349 and JP23K13149.
MJJ and SC acknowledge support for the current research from the National Research Foundation (NRF) of Korea under the programs 2022R1A2C1003130, RS-2023-00219959, and RS-2024-00413036.
Support was provided to JDRP through program HST-GO-16264. JDRP is supported by NASA through an Einstein Fellowship grant No.~HF2-51541.001, awarded by STScI, which is operated by the Association of Universities for Research in Astronomy, Inc., for NASA, under contract NAS5-26555.
AZ acknowledges support by Grant No.~2020750 from the United States-Israel Binational Science Foundation (BSF) and Grant No.~2109066 from the United States National Science Foundation (NSF), and by the Israel Science Foundation Grant No. 864/23.

\end{acknowledgements}

\bibliographystyle{aa}
\bibliography{references}


\begin{appendix}

\begin{landscape}
\section{Measured photometry of cluster members in MACS J0138$-$2155}
\label{app:photometry}

In Table \ref{tab:mag_cat}, we list the photometric ID, the position in RA and dec, and the magnitudes in all filters listed in Table~\ref{tab:encore_img_data} of the objects included in the model. The full table is available at the CDS. The BCG is labeled as index 0. If an object is not modeled in a specific band (e.g. if it is not in the field of view) its magnitude is marked as $-$. The reported magnitudes are AB magnitudes.

\begin{table}[H]
\caption{Photometry of cluster members in MACS J0138$-$2155.}
\centering
\begin{tabularx}{\linewidth}{p{1cm} p{1.8cm} p{1.8cm} *{11}{p{1.2cm}}}\toprule \toprule
        ID  & RA & dec & \multicolumn{11}{c}{Magnitudes} \\ 
        \cmidrule(lr){4-14}
        
        & & & $m_{\rm F555W}$ & $m_{\rm F814W}$ & $m_{\rm F105W}$  & $m_{\rm F115W}$ & $m_{\rm F125W}$ & $m_{\rm F150W}$ & $m_{\rm F160W}$ & $m_{\rm F200W}$  & $m_{\rm F277W}$ & $m_{\rm F356W}$ & $m_{\rm F444W}$ \\ \midrule \midrule
                
0 & 24.5157084 & $-$21.9254717 & 17.57 & 15.11 & 15.84 & 15.85 & 15.56 & 15.47 & 15.30 & 15.22 & 15.08 & 15.51 & 15.60 \\
1 & 24.5195585 & $-$21.9489923 & 21.41 & 19.95 & 19.28 & 19.02 & 19.06 & 18.78 & 18.71 & 18.53 & 18.62 & 19.12 & 19.37 \\
3 & 24.5133872 & $-$21.9453005 & 21.72 & 20.31 & 19.82 & 19.56 & 19.58 & 19.22 & 19.32 & 19.04 & 19.22 & 19.74 & 19.99 \\
9 & 24.5148245 & $-$21.9415458 & 23.28 & 21.90 & 21.34 & 21.11 & 21.16 & 20.81 & 20.91 & 20.63 & 20.76 & 21.29 & 21.61 \\
12 & 24.5268991 & $-$21.9405099 & 22.89 & 21.61 & 21.15 & 20.91 & 20.93 & 20.55 & 20.7 & 20.37 & 20.59 & 21.14 & 21.36 \\
13 & 24.5139526 & $-$21.9411666 & 23.30 & 21.90 & 21.36 & 20.96 & 21.25 & 20.75 & 20.98 & 20.58 & 20.74 & 21.32 & 21.46 \\
14 & 24.5187763 & $-$21.9396492 & 21.35 & 20.04 & 19.53 & 19.34 & 19.30 & 19.03 & 19.02 & 18.81 & 18.94 & 19.47 & 19.67  \\
17 & 24.514007 & $-$21.9402814 & 21.22 & 19.82 & 19.36 & 19.16 & 19.15 & 18.82 & 18.87 & 18.61 & 18.69 & 19.19 & 19.40  \\
23 & 24.5261825 & $-$21.9377475 & 22.26 & 21.01 & 20.35 & 20.23 & 20.16 & 19.96 & 19.68 & 19.83 & 20.01 & 20.53 & 20.76  \\
. & . & . & . & . & . & . & . & . & . & . & . & . & .  \\
. & . & . & . & . & . & . & . & . & . & . & . & . & .  \\

             \bottomrule  

\end{tabularx}
        \label{tab:mag_cat}
\end{table}
\end{landscape}

\section{Lens models}
\label{app:models}
In Table \ref{tab:mass_models}, we provide an overview of the lens mass models. We describe the different models and list their number of mass parameters ($N_{\rm par}$), as well as their $\chi_{\rm im,MP}^2$ values.

\begin{table*}
    \centering
    \caption{Lens mass models.}
    \label{tab:model_fits}
    \begin{tabular}{lp{7.5cm}ccc}
        \hline \addlinespace[2pt]
        Model Name & Description & \multicolumn{2}{c}{Number of mass parameters $N_{\rm par}$} & $\chi_{\rm im,MP}^2$  \\  \addlinespace[2pt]
        \cline{3-4} \addlinespace[2pt]
                            &                      &  variable & with Gaussian priors &  \\ \hline \addlinespace[2pt]
        \texttt{iso\_halo}             &  & 25 & 4 & 6.3 \\
        \addlinespace[2pt]
         & $\bullet$ primary cluster DM halo: isothermal elliptical mass distribution   & \hspace{0.5cm} 6 & \hspace{0.5cm} 0 &  \\
         & $\bullet$ perturbative cluster DM halo: isothermal elliptical mass distribution & \hspace{0.5cm} 6 & \hspace{0.5cm} 0 & \\
         & $\bullet$ cluster members scaling relation for $\theta_{\rm E,\infty}^{\rm ref}$ and $r_{\rm t}^{\rm ref}$, with Gaussian prior on velocity dispersion based on Faber-Jackson & \hspace{0.5cm} 2 & \hspace{0.5cm} 1 & \\
         & $\bullet$ circular BCG with Gaussian prior on $\theta_{\rm E,\infty,BCG}$ from velocity-dispersion measurement and uniform prior on $r_{\rm t,BCG}$ & \hspace{0.5cm} 2 & \hspace{0.5cm} 1 & \\
         & $\bullet$ cluster member ID$_{\rm phot}$ 116 with Gaussian prior on $\theta_{\rm E,\infty,116}$ from velocity-dispersion measurement and uniform prior on $r_{\rm t,116}$ & \hspace{0.5cm} 2 & \hspace{0.5cm} 1 & \\
         & $\bullet$ three jellyfish galaxies with variable Einstein radii & \hspace{0.5cm} 3 & \hspace{0.5cm} 0 & \\
         & $\bullet$ foreground LOS galaxy at $z_{\rm fg}=0.309$ with variable $\theta_{\rm E,\infty,fg}$ and $r_{\rm t,fg}$  & \hspace{0.5cm} 2 & \hspace{0.5cm} 0 & \\
         & $\bullet$ background LOS galaxy at $z_{\rm bg}=0.371$ with  $\theta_{\rm E,\infty,bg}$ having prior from velocity dispersion and variable $r_{\rm t,bg}$ with uniform prior & \hspace{0.5cm} 2 & \hspace{0.5cm} 1 & \\ \addlinespace[2pt]
        \hline \addlinespace[2pt]
        \texttt{PL\_halo} & & 26 & 4 & 6.3 \\
        & $\bullet$ primary cluster DM halo: power-law elliptical mass distribution             & \hspace{0.5cm} 7 & \hspace{0.5cm} 0 &  \\ 
        & $\bullet$ other components are the same as the {\texttt{iso\_halo}} model above             & \hspace{0.5cm} 19 & \hspace{0.5cm}  4 &  \\ 
        \addlinespace[2pt]
        \hline \addlinespace[2pt]
        \texttt{NFW\_halo} & & 25 & 4 & 23 \\ 
        & $\bullet$ primary cluster DM halo: elliptical NFW mass distribution             & \hspace{0.5cm} 6 & \hspace{0.5cm} 0 &  \\ 
        & $\bullet$ other components are the same as the {\texttt{iso\_halo}} model above             & \hspace{0.5cm} 19 & \hspace{0.5cm} 4 &  \\ 
        \addlinespace[2pt]
        \hline \addlinespace[2pt]
        
        \texttt{iso\_halo+sheet}             &  & 26 & 4 & 6.3 \\       
        & $\bullet$ same components as the {\texttt{iso\_halo}} model above             & \hspace{0.5cm} 25 & \hspace{0.5cm} 4 &  \\ 
        & $\bullet$ mass-sheet parameter             & \hspace{0.5cm} 1 & \hspace{0.5cm} 0 & \\ 
        \addlinespace[2pt]
        \hline \addlinespace[2pt]
        
        \texttt{iso\_halo+ell\_BCG}             &  & 27 & 4 & 5.7 \\       
        & $\bullet$ same components as the {\texttt{iso\_halo}} model, excluding the BCG (elliptical instead of circular)            & \hspace{0.5cm} 23 & \hspace{0.5cm} 3 &  \\ 
        & $\bullet$ elliptical BCG with Gaussian prior on $\theta_{\rm E,\infty,BCG}$ from velocity-dispersion measurement and uniform priors on $r_{\rm t,BCG}$, $q_{{\rm BCG}}$ and $\phi_{{\rm BCG}}$    & \hspace{0.5cm} 4 & \hspace{0.5cm} 1 & \\ 
        \addlinespace[2pt]
        \hline \addlinespace[2pt]        

        \texttt{single\_plane}             &  & 25 & 4 & 6.1 \\       
        & $\bullet$ same components as the {\texttt{iso\_halo}} model above except the foreground galaxy (at $z_{\rm fg}=0.309$) and background galaxy (at $z_{\rm bg}=0.371$) are assumed to be at the cluster redshift of $z_{\rm d}=0.336$.  The positional uncertainties of the multiple image positions are elliptical, as listed in Table \ref{tab:imagepos} and used in the above models.              & \hspace{0.5cm} 25 & \hspace{0.5cm} 4 &  \\ 
        \addlinespace[2pt]
        \hline \addlinespace[2pt]

        \texttt{baseline}             &  & 25 & 4 & 7.3 \\       
        & $\bullet$ same components as the {\texttt{single\_plane}} model above except the positional uncertainties of the multiple image positions are circularized from the elliptical uncertainties.  The circularized uncertainty is taken to be the geometric mean of the semi-major and semi-minor axis of the elliptical uncertainty ($\sigma^{\rm circ} = \sqrt{\sigma^{\rm major} \sigma^{\rm minor}}$).            & \hspace{0.5cm} 25 & \hspace{0.5cm} 4 &  \\ 
        \addlinespace[2pt]
        \hline \addlinespace[2pt]    
    \end{tabular}
\caption*{Notes: The first line of each model lists the total number of parameters, whereas the subsequent lines indicate the numbers of parameters for the specific components of the mass model. The last column ($\chi^2_{\rm im,MP}$) is the image-position $\chi^2$ of the most-probable mass model.  The galaxy mass centroids are located at their observed light centroids, and the galaxies are assumed to be circular, except for the BCG where we considered both circular and elliptical mass distributions.  Therefore, all galaxies following the scaling relations are described by two parameters ($\theta_{\rm E,\infty}^{\rm ref}$ and $r_{\rm t}^{\rm ref}$), whereas galaxies not following the scaling relations have each two parameters ($\theta_{{\rm E,\infty},i}$ and $r_{{\rm t},i}$ for galaxy $i$) when they are circular, and four parameters ($\theta_{{\rm E,\infty,BCG}}$, $r_{{\rm t,BCG}}$, $q_{{\rm BCG}}$ and $\phi_{{\rm BCG}}$)  when elliptical.  The number of free parameters for the DM halo parametrizations is described in Sect.~\ref{sec:mass_model:parametrizations:clusterhalo} and \ref{sec:mass_model:parametrizations:sheet} and listed in the table. }
    \label{tab:mass_models}
\end{table*}

\begin{landscape}

\section{Model parameters}
\label{app:parameters}
In Table \ref{tab:encore_final_param}, we list the final model parameter values for the four multi-plane lens models (\texttt{iso\_halo}, \texttt{PL\_halo}, \texttt{iso\_halo+sheet} and \texttt{iso\_halo+ell\_BCG}) and the single-plane \texttt{baseline} model.

\begin{table}[H]
\centering
\fontsize{9}{9}\selectfont
\caption[SN Encore final model parameter values.]{Model parameter priors and inferred values from observed image positions and stellar velocity dispersion measurements. }
\begin{tabular}{lllccccc} \toprule
Parameter & Description & Prior & \texttt{iso\_halo} & \texttt{PL\_halo} & \texttt{iso\_halo+sheet} & \texttt{iso\_halo+ell\_BCG} & \texttt{baseline} \\ \midrule \midrule
                     \multicolumn{8}{l}{Scaling relation (ID$_{\rm phot}=115$)} \\
                     \midrule
                  $\sigma_{\rm v,115}\ [\textrm{km/s}]$   & velocity dispersion & $\mathcal{G}$(206,25) & $180_{-22}^{+21}$ & $180\pm21$ & $170\pm21$   & $180_{-22}^{+20}$ & $170_{-22}^{+23}$ 
                  \vspace{2px}\\  
                  $r_{\rm t,115}\ [\arcsec] $   & truncation radius & flat(1.8,40) & $2.2_{-0.3}^{+0.9}$ & $2.4_{-0.5}^{+1.1}$ & $2.2_{-0.3}^{+0.7}$ & $2.3_{-0.4}^{+0.8}$ & $2.2_{-0.3}^{+0.6}$   \vspace{2px} \\  
                      \midrule
                  \multicolumn{8}{l}{BCG} \\
                     \midrule
                  $(x_{\rm BCG}, y_{\rm BCG})\ [\arcsec]$   & centroid & exact & (0, 0) & (0, 0)  & (0, 0) & (0, 0) & (0, 0)    \\
                  $q_{\rm BCG}$   & axis ratio & exact or flat(0.4,1) & 1 & 1 & 1 &  $0.89_{-0.12}^{+0.08}$ & 1 \\
                  $\phi_{\rm BCG}\ [^{\circ}]$   & position angle & flat(0,360) &  ---  & --- & --- & $36 \pm 63$ & --- \\
                  $\theta_{\rm E,\infty,BCG}\ [\arcsec]$   & Einstein radius & $\mathcal{G}$(3.08,0.32)$^{\dagger}$ & $3.36_{-0.39}^{+0.25}$  & $3.17_{-0.23}^{+0.30}$   & $3.07_{-0.25}^{+0.34}$ & $2.99_{-0.22}^{+0.24}$ & $3.16_{-0.24}^{+0.29}$  \vspace{2px}\\  
                  $r_{\rm t,BCG}\ [\arcsec] $   & truncation radius & flat(1,50) & $41_{-14}^{+13}$  & $35_{-20}^{+18}$ & $33_{-12}^{+16}$ & $27_{-12}^{+15}$ & $33_{-13}^{+18}$ \vspace{2px}  \\
                  
                 \midrule
                 \multicolumn{8}{l}{Cluster member not in scaling relation} \\
                     \midrule
                     $(x_{\rm 116}, y_{\rm 116})\ [\arcsec]$   & centroid & exact & (0.314, 9.819) & (0.314, 9.819) & (0.314, 9.819)  & (0.314, 9.819) & (0.314, 9.819)    \vspace{2px} \\
                    $q_{\rm 116}$   & axis ratio & exact & 1 & 1  & 1 & 1 & 1 \\
                    $\theta_{\rm E,\infty,116}\ [\arcsec]$   & Einstein radius G1 & $\mathcal{G}$(1.67,0.19)$^{\dagger}$ & $1.6\pm0.2$ & $1.6\pm0.2$ & $1.6\pm0.2$ & $1.6\pm0.2$  & $1.6\pm0.2$ \vspace{2px} \\
                    $r_{\rm t,116}\ [\arcsec]$   & truncation radius &flat(0.05,10)  & $1.1_{-0.4}^{+0.5}$ &  $1.1\pm0.4$  &  $1.1_{-0.4}^{+0.5}$ & $1.1_{-0.3}^{+0.4}$  &  $1.0_{-0.4}^{+0.5}$ \vspace{2px} \\

                  \midrule
                     \multicolumn{8}{l}{Jellyfish galaxies} \\
                     \midrule
                     $(x_{\rm JF1}, y_{\rm JF1})\ [\arcsec]$   & centroid & exact & (19.094, $-$13.361) & (19.094, $-$13.361) & (19.094, $-$13.361) & (19.094, $-$13.361) & (19.094, $-$13.361)   \vspace{2px} \\
                   $\theta_{\rm E,\infty,JF1}\ [\arcsec] $   & Einstein radius & flat(0.01,10) & $0.25\pm0.07$  & $0.25\pm0.08$  & $0.23_{-0.10}^{+0.09}$ & $0.20_{-0.06}^{+0.05}$ & $0.13_{-0.05}^{+0.06}$ \vspace{2px} \\
                   $r_{\rm t,JF1}\ [\arcsec]$   & truncation radius & exact  & 5 & 5 & 5 & 5 & 5 \vspace{2px}  \\
                   & & & \\
                    $(x_{\rm JF2}, y_{\rm JF2})\ [\arcsec]$   & centroid & exact & ($-$5.114, 6.941) & ($-$5.114, 6.941) & ($-$5.114, 6.941) & ($-$5.114, 6.941) & ($-$5.114, 6.941)  \vspace{2px}  \\
                   $\theta_{\rm E,\infty,JF2}\ [\arcsec] $   & Einstein radius &flat(0.01,10) & $0.79_{-0.14}^{+0.16}$ & $0.64_{-0.14}^{+0.12}$ & $0.63\pm0.14$ & $0.73_{-0.16}^{+0.17}$ & $0.57_{-0.24}^{+0.27}$ \vspace{2px}  \\
                   $r_{\rm t,JF2}\ [\arcsec]$   & truncation radius & exact  &  5  & 5 & 5 & 5 & 5 \vspace{2px} \\
                   & & & \\
                    $(x_{\rm JF3}, y_{\rm JF3})\ [\arcsec]$   & centroid & exact & ($-$1.649, 2.597) & ($-$1.649, 2.597) & ($-$1.649, 2.597) & ($-$1.649, 2.597) & ($-$1.649, 2.597) \vspace{2px}  \\
                   $\theta_{\rm E,\infty,JF3}\ [\arcsec] $   & Einstein radius &flat(0.01,10) & $1.5_{-0.6}^{+0.4}$ & $1.3_{-0.4}^{+0.5}$  & $1.3_{-0.3}^{+0.4}$  & $0.9_{-0.5}^{+0.4}$ & $1.4_{-0.3}^{+0.4}$   \vspace{2px}\\
                   $r_{\rm t,JF3}\ [\arcsec]$   & truncation radius & exact  & 5  & 5  & 5 & 5 & 5 \vspace{2px} \\
               
             \bottomrule
              \vspace{2px}
             \end{tabular}
        \label{tab:encore_final_param}

           \caption*{Notes. Columns 1 and 2 are the parameter and description.  Column 3 is the prior on the parameter, where the letter $\mathcal{G}$ denotes a Gaussian prior, with its center and $1\sigma$ width in parentheses. The Gaussian priors on the Einstein radii with a $^{\dagger}$ symbol are based on the first version of velocity dispersion measurements from \citet{Granata+24} using an older version of stellar template libraries; the final velocity dispersion measurements change by $\lesssim$1\,km\,s$^{-1}$, which has negligible effect on the modeling results. Columns 4 to 7 list the inferred parameter values of the corresponding mass models indicated in the header line.  Parameter values that have no uncertainties are parameters that are fixed with priors listed as `exact'.  A dashed line in the parameter value indicates that the parameter is not present in the particular mass model. Position angles are measured east of north in degrees.} 
\end{table}
\end{landscape}

\addtocounter{table}{-1}

\begin{landscape}
\begin{table}
\centering
\fontsize{9}{9}\selectfont
\caption{Continued.}
\begin{tabular}{lllccccc} \toprule
Parameter & Description & Prior & \texttt{iso\_halo} & \texttt{PL\_halo} & \texttt{iso\_halo+sheet} & \texttt{iso\_halo+ell\_BCG} & \texttt{baseline} \\ \midrule \midrule

                  \multicolumn{8}{l}{Foreground galaxy} \\
                     \midrule
                  $(x_{\rm fg}, y_{\rm fg})\ [\arcsec]$   & centroid & exact & ($-$0.868, $-$16.564)  & ($-$0.868, $-$16.564) & ($-$0.868, $-$16.564) & ($-$0.868, $-$16.564) & ($-$0.868, $-$16.564) \vspace{2px}  \\
                  $\theta_{\rm E,\infty,fg}\ [\arcsec]$   & Einstein radius & flat(0.1,10) & $0.20_{-0.03}^{+0.02}$ & $0.24\pm0.04$  & $0.21_{-0.03}^{+0.05}$ & $0.20\pm0.03$ & $0.18_{-0.02}^{+0.04}$ \vspace{2px} \\  
                  $r_{\rm t,fg}\ [\arcsec] $   & truncation radius & flat(0.1,40) & $16_{-11}^{+14}$ & $2.3_{-0.9}^{+1.5}$ & $7_{-4}^{+13}$ & $18_{-14}^{+15}$ & $8_{-7}^{+28}$
                  \vspace{2px} \\  

                     \midrule
                  \multicolumn{8}{l}{Background galaxy} \\
                     \midrule
                  $(x_{\rm bkg}, y_{\rm bkg})\ [\arcsec]$   & centroid & exact & (6.871, 3.843)  & (6.871, 3.843) & (6.871, 3.843) & (6.871, 3.843)  & (6.871, 3.843) \vspace{2px} \\
                  $\theta_{\rm E,\infty,bkg}\ [\arcsec]$   & Einstein radius & $\mathcal{G}$(1.66,0.10)$^{\dagger}$ & $1.65_{-0.09}^{+0.10}$  &  $1.67_{-0.12}^{+0.11}$ &  $1.67_{-0.09}^{+0.10}$  & $1.67\pm0.09$ & $1.64\pm0.10$ \vspace{2px} \\  
                  $r_{\rm t,bkg}\ [\arcsec] $   & truncation radius & flat(0.1,40) & $26_{-14}^{+10}$ & $22\pm12$ & $15_{-8}^{+20}$ &  $14_{-8}^{+14}$ & $14_{-8}^{+18}$ \vspace{2px} \\  

                     \midrule
                    \multicolumn{8}{l}{Dark matter halos} \\
                     \midrule
                    $x_{\rm DM1}\ [\arcsec]$   & $x$ centroid & flat($-$20,20) & $-0.1\pm0.3$     & $-0.2\pm0.3$ & $-0.1\pm0.3$ & $-0.3\pm0.3$ & $0.1\pm0.3$ \vspace{2px} \\
                    $y_{\rm DM1}\ [\arcsec]$   & $y$ centroid & flat($-$20,20) & $-1.2\pm0.5$     & $-0.9_{-0.5}^{+0.4}$ & $-1.1_{-0.4}^{+0.3}$ & $-0.6_{-0.4}^{+0.5}$ & $-1.1\pm0.4$ \vspace{2px} \\
                    $q_{\rm DM1}$   & axis ratio &flat(0.2,1) & $0.64\pm0.02$ & $0.64\pm0.02$   & $0.63\pm0.02$  & $0.66_{-0.02}^{+0.03}$ & $0.63\pm0.02$ \vspace{2px} \\
                    $\phi_{\rm DM1}\ [^{\circ}]$   & position angle & flat(0,360)& $43\pm1$ & $43_{-2}^{+1}$  & $44\pm2$ & $43_{-1}^{+2}$  & $45\pm2$  \vspace{2px}\\
                    $\theta_{\rm E,\infty,DM1}\ [\arcsec]$   & Einstein radius  & flat(0,50) & $20.7_{-1.3}^{+1.1}$ & --- & $20.2_{-1.3}^{+1.4}$   & $19.5_{-1.2}^{+1.8}$  & $21.4_{-1.9}^{+1.6}$  \vspace{2px} \\
                    $E\ [\arcsec]$   & halo strength  & flat(0,50) & --- &  $13\pm2$ & --- & --- & --- \vspace{2px} \\
                    $r_{\rm c,DM1}\ [\arcsec]$   & core radius &flat(0,50) & $6.5_{-0.8}^{+0.7}$ & $6.4_{-0.8}^{+0.6}$ & $6.1_{-0.7}^{+0.6}$ & $5.5_{-0.5}^{+0.6}$ & $6.6\pm0.8$   \vspace{2px} \\
                    $\gamma$   & radial slope & flat(0.,2.) & --- & $0.51\pm0.03$  & --- & --- & --- \vspace{2px} \\
                    & & & & \\
                    $x_{\rm DM2}\ [\arcsec]$   & $x$ centroid & flat($-$30,30) & $25_{-5}^{+4}$ & $20_{-5}^{+6}$ & $21_{-8}^{+6}$  & $18_{-3}^{+8}$ & $19_{-6}^{+9}$   \vspace{2px} \\
                    $y_{\rm DM2}\ [\arcsec]$   & $y$ centroid & flat($-$30,30) & $-16_{-7}^{+10}$ & $-19_{-6}^{+7}$ & $-20_{-7}^{+15}$  & $-22\pm4$ & $-20_{-7}^{+10}$   \vspace{2px} \\
                    $q_{\rm DM2}$   & axis ratio  &flat(0.2,1) & $0.38_{-0.08}^{+0.10}$ & $0.50_{-0.14}^{+0.08}$  & $0.35_{-0.10}^{+0.15}$ & $0.45_{-0.14}^{+0.12}$ & $0.50_{-0.11}^{+0.09}$ \vspace{2px}  \\
                    $\phi_{\rm DM2}\ [^{\circ}]$   & position angle & flat(0,360)& $147_{-5}^{+4}$ & $145\pm3$ & $148_{-4}^{+5}$ & $141\pm4$ & $148_{-4}^{+3}$ \vspace{2px} \\
                    $\theta_{\rm E,\infty,DM2}\ [\arcsec]$   & Einstein radius & flat(0,50) & $20_{-6}^{+7}$ & $21_{-5}^{+6}$ & $15_{-6}^{+13}$  &  $23\pm6$ & $25_{-6}^{+8}$ \vspace{2px} \\
                    $r_{\rm c,DM2}\ [\arcsec]$   & core radius &flat(0,70) & $27_{-7}^{+6}$  & $29_{-6}^{+8}$  & $32_{-9}^{+11}$ & $24\pm5$ & $50_{-21}^{+16}$  \vspace{2px} \\   
                    
                     \midrule
                    \multicolumn{8}{l}{Mass sheet} \\
                     \midrule
                    $\kappa_0$   & mass sheet & flat($-1$,1) & ---  & --- & $0.09_{-0.12}^{+0.07} $ & --- & --- \vspace{2px} \\

             \bottomrule
             \end{tabular}
        \label{tab:encore_final_param2}

\end{table}
\end{landscape}

\section{Predicted image positions and magnifications of supernovae from the \texttt{baseline} model}
\label{app:SNpred:baseline}

In Table \ref{tab:SNpred:baseline} we list the predicted image positions and magnifications of SN Encore and SN Requiem from the \texttt{baseline} model where the LOS galaxies are approximated to be at the cluster redshift plane, and positional uncertainties of lensed images are circularized (see Sect.~\ref{sec:results_fixed_cosmo:baseline} for more details).  Comparing to the ultimate model predictions in Table \ref{tab:SNpred:best} from multi-plane lens modelling, the predictions from the \texttt{baseline} model are mostly consistent within the 1$\sigma$ modelling uncertainties.  This shows that the \texttt{baseline} model provides a good approximation to the final best model for the current sets of image positions and kinematic data.

\begin{table}[h!]
    \centering
    \caption{Predicted image positions and magnifications of SN Encore and SN Requiem from the \texttt{baseline} lens mass model.}
    \begin{tabular}{lccc}
        \toprule
        image & $x$ coordinate $[\arcsec]$ & $y$ coordinate $[\arcsec]$ & magnification \\ 
        \midrule
        \midrule
        \multicolumn{4}{l}{SN Encore predictions (\texttt{baseline} model)} \\
        \midrule
         1a & $\phantom{-}8.41\pm{0.04}$ & $-16.07\pm{0.01}$ & $-26.6^{+2.5}_{-2.7}$ \vspace{2px} \\
         1b & $\phantom{-}0.22\pm{0.04}$  & $-17.86\pm{0.01}$ & $\phantom{-}39.6^{+5.9}_{-6.1}$ \vspace{2px} \\
         1c* & $\phantom{-}19.48^{+0.03}_{-0.04}$ & $-2.64^{+0.10}_{-0.09}$ & $\phantom{-}11.0^{+0.8}_{-0.9}$ \vspace{2px}\\
         1d* & $-6.14^{+0.09}_{-0.10}$ & $\phantom{-}7.66^{+0.07}_{-0.09}$ & $-2.6^{+0.5}_{-0.7}$ \\
         1e*$^{\dagger}$ & $-1.20^{+0.27}_{-0.41}$ & $1.70^{+0.71}_{-0.41}$ & $\phantom{-}5.3^{+4.6}_{-2.4}$ \\
        \midrule
        \multicolumn{4}{l}{SN Requiem predictions (\texttt{baseline} model)} \\
        \midrule
        2a & $11.29\pm{0.07}$ & $-15.49\pm{0.05}$       & $-33.8^{+3.5}_{-4.2}$ \vspace{2px} \\
        2b & $2.01^{+0.08}_{-0.07}$ & $-18.62\pm{0.02}$ & $\phantom{-}23.8\pm{2.3}$ \vspace{2px} \\
        2c & $18.86\pm{0.03}$ & $-6.64\pm{0.08}$ & $\phantom{-}14.9^{+1.7}_{-1.6}$ \vspace{2px} \\
        2d* & $-4.23^{+0.28}_{-0.25}$ & $5.80^{+0.37}_{-0.39}$ & $-2.3^{+0.4}_{-0.7}$ \vspace{2px} \\
        \bottomrule
    \end{tabular}
    \label{tab:SNpred:baseline}
    \tablefoot{Notes. The coordinates are relative to the BCG that is centred at $(x,y)=(0,0)$.  Lensed images marked by * are model predictions which have not yet been clearly detected, either due to their faintness (image 1c) or due to their long time delays (images 1d, 1e and 2d). Lensed image 1e marked by $^{\dagger}$ is predicted by 13\% of the models in the \texttt{baseline} MCMC chain; all the other images listed above are predicted by all the models in the chain.  Image 2e is not predicted in the \texttt{baseline} model.    }
\end{table}

\end{appendix}

\end{document}